\documentclass[twocolumn, prl, amssymb, superscriptaddress, aps, showpacs,preprintnumbers,
amsmath,showkeys,floatfix]{revtex4-2}

\usepackage{graphics}
\usepackage{graphicx}
\usepackage{dcolumn}
\usepackage{bm}
\usepackage{longtable}
\usepackage{epsfig}
\usepackage{times}
\usepackage{url}
\usepackage{xcolor, soul}

\begin{document}

	\title{Program-Synthesis-Driven Autodesign of Universal Unitary Operators}
	
	\author{Yifei \surname{Zhang}}
	\thanks{These authors contributed equally to this work.}
	\affiliation{Key Laboratory of Nuclear Physics and Ion-beam Application (MOE), Institute of Modern Physics, Fudan University, Shanghai 200433, China}
	\affiliation{Research Center for Theoretical Nuclear Physics, NSFC and Fudan University, Shanghai 200438, China}
	
	\author{Dong \surname{Chen}}
	\thanks{These authors contributed equally to this work.}
	\affiliation{Huawei Technologies Co., Ltd, Beijing 100095, China}
	
	\author{Fan \surname{Wang}}
	\affiliation{Department of Industrial Engineering and Decision Analytics, Hong Kong University of Science and Technology, HongKong, China}
	\affiliation{Research Center for Theoretical Nuclear Physics, NSFC and Fudan University, Shanghai 200438, China}
	
	\author{Wenrui \surname{Zhang}}
	\affiliation{Huawei Technologies Co., Ltd, Beijing 100095, China}
	
	\author{Yan \surname{Chen}}
	\affiliation{Hunan Key Laboratory of Mechanism and Technology of Quantum Information, Changsha 410073, China}
	
	\author{Dingding \surname{Han}}
	\affiliation{School of Information Science and Technology, Fudan University, Shanghai 200433, China}
	\affiliation{Research Institute of Intelligent Complex Systems, Fudan University, Shanghai 200433, China}
	
	\author{Jianmin \surname{Yuan}}
	\affiliation{Institute of Atomic and Molecular Physics, Jilin University, Changchun 130012, China}
	
	\author{Xiangjin \surname{Kong}}
	\email{kongxiangjin@fudan.edu.cn}
	\affiliation{Key Laboratory of Nuclear Physics and Ion-beam Application (MOE), Institute of Modern Physics, Fudan University, Shanghai 200433, China}
	\affiliation{Research Center for Theoretical Nuclear Physics, NSFC and Fudan University, Shanghai 200438, China}
	
	\author{Yu-Gang \surname{Ma}}
	\email{mayugang@fudan.edu.cn}
	\affiliation{Key Laboratory of Nuclear Physics and Ion-beam Application (MOE), Institute of Modern Physics, Fudan University, Shanghai 200433, China}
	\affiliation{Research Center for Theoretical Nuclear Physics, NSFC and Fudan University, Shanghai 200438, China}
	\affiliation{School of Physics, East China Normal University, Shanghai 200062, China}

	\begin{abstract}
		
		We demonstrate that AI-driven program synthesis can autonomously discover fundamental strategies for decomposing unitary matrices in photonic networks. By extending DreamCoder to complex-valued linear algebra, the system generates decomposition programs achieving the minimal $N(N-1)/2$ Mach--Zehnder interferometers, distinct from both Reck and Clements architectures. Learned programs encode dimension-agnostic invariants: strategies discovered for $5 \times 5$ matrices generalize to higher dimensions such as $64 \times 64$. The discovered programs encode interpretable, dimension-agnostic construction rules. These rules generalize across matrix sizes without retraining, demonstrating that autonomous program synthesis can serve as a scalable paradigm for algorithm discovery and the automated design of universal unitary operators. Beyond universal decompositions, the system automatically exploits matrix structure to reduce the interferometer count below the universal theoretical bound. For instance, for Householder matrices, it discovers a dimension-independent rule that requires only $2N - 3$ MZIs. This achieves linear, rather than quadratic, scaling and generalizes to arbitrary $N$ without retraining. For matrices obtained from the singular value decomposition of sparse matrices, reductions generally increase with sparsity, reaching up to $38\%$ fewer MZIs than the universal theoretical bound $N(N-1)/2$ at $95\%$ sparsity. These MZI reductions translate directly into practical hardware benefits for scalable photonic implementations. Taken together, the system functions as a single unified engine that discovers both universal decomposition rules and matrix-specific optimizations, without being provided with the structural or analytical properties of the input matrices.
		
	\end{abstract}
	
	\maketitle
	
	Unitary transformations lie at the foundation of quantum mechanics, governing the evolution of quantum states in closed systems~\cite{nielsen2010quantum}, and are central to modern photonic computing and quantum information processing~\cite{kok2007linear,shastri2021photonics,mcmahon2023physics}. In linear optical systems, unitary matrices determine how multimode fields interfere and propagate through reconfigurable interferometric networks, allowing direct hardware implementations of matrix operations with ultrahigh speed, low energy consumption, and intrinsic parallelism~\cite{knill2001scheme,mcmahon2023physics}. These capabilities make photonic platforms attractive for optical neural networks~\cite{shen2017deep,perez2017multipurpose,lin2018all,pai2023experimentally,xue2024fully,fu2024optical}, where unitaries implement trainable linear layers, and for quantum computation~\cite{bogaerts2020programmable,kok2007linear,o2007optical,konno2024logical,psiquantum2025manufacturable}, where they realize universal linear-optical circuits. Rapid advances in integrated photonics have produced large-scale, low-loss, and programmable interferometer meshes~\cite{pelucchi2022potential,bente2025potential,shekhar2024roadmapping}, motivating scalable and systematic methods to realize arbitrary high-dimensional unitaries in hardware.
	
	In practice, arbitrary $N \times N$ unitary transformations are implemented by decomposing them into sequences of Mach--Zehnder interferometers (MZIs), which serve as universal two-dimensional beam-splitter gates~\cite{miller2013self,carolan2015universal}. Mathematically, any matrix can be expressed via singular value decomposition (SVD) as $W = U \Sigma V^\dagger$~\cite{golub1965calculating}, producing two unitary matrices $U$ and $V^\dagger$ and a diagonal matrix $\Sigma$. Since diagonal matrices $\Sigma$ can be directly realized with MZIs, the challenge reduces to efficiently decomposing the unitary components $U$ and $V^\dagger$. Classical schemes, such as the Reck triangular mesh~\cite{reck1994experimental} and the Clements rectangular mesh~\cite{clements2016optimal}, provide hand-derived constructions using $N(N-1)/2$ MZIs, forming the algorithmic backbone for universal linear optics and underpinning most present-day photonic processors~\cite{bogaerts2020programmable,perez2017multipurpose}.
	
	\begin{figure*}[!t]
		\centering
		\begin{minipage}[t]{0.48\textwidth}
			\vspace{0pt}
			\llap{\textbf{(a)}\hspace{0.3em}}%
			\includegraphics[width=\textwidth]{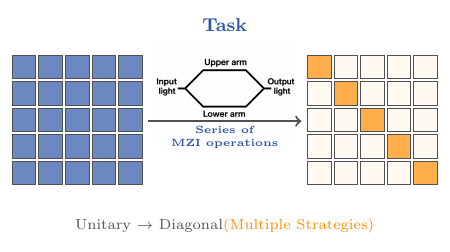}
		\end{minipage}
		\hfill
		\begin{minipage}[t]{0.42\textwidth}
			\vspace{0pt}
			\llap{\textbf{(b)}\hspace{0.3em}}%
			\includegraphics[width=\textwidth]{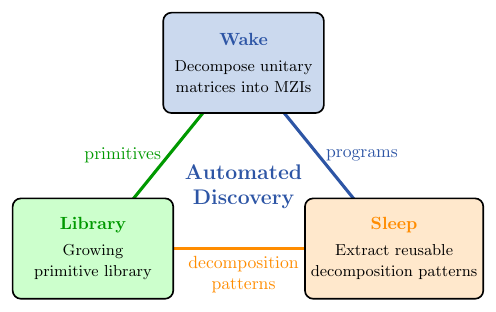}
		\end{minipage}
		\caption{
			Autonomous discovery of unitary decomposition strategies.
			(a) Task transformation: unitary matrices are decomposed into diagonal form through a series of MZI operations, with multiple decomposition strategies available.
			(b) DreamCoder's abstraction loop: candidate MZI programs are enumerated and tested on decomposition tasks, recurring fragments from successful programs are compressed into reusable library primitives, and the enlarged library is reused to bias subsequent searches toward deeper decompositions.
		}
		\label{fig:main_results}
	\end{figure*}
	
	Despite their foundational role, these schemes reflect specific human-designed elimination strategies derived from analytic reasoning and occupy only a small region of the combinatorial space of possible elimination orderings, leaving open the question of whether alternative, equally fundamental strategies exist. To address this, we employ automated reasoning to systematically explore this landscape, uncovering strategies that extend beyond human intuition and yield generalizable constructions. In pursuit of this goal, we develop a synthesis engine—built on the DreamCoder framework~\cite{ellis2021dreamcoder} and extended to manipulate complex-valued linear algebra—that autonomously discovers strategies for decomposing unitary matrices. The synthesized programs are not merely confined to known triangular or rectangular meshes. Instead, the system identifies previously unreported elimination orderings that achieve the optimal $N(N-1)/2$ MZIs. Analysis of the learned programs reveals that these constructions encode dimension-agnostic algorithmic regularities: strategies discovered on $5 \times 5$ matrices generalize to arbitrary higher dimensions such as $64 \times 64$ without retraining, indicating that the system has internalized domain-general patterns rather than dimension-specific heuristics. These results establish program synthesis as a means of autonomously identifying algorithmic structure beyond known human-designed schemes, enabling the discovery of interpretable and generalizable rules for universal unitary decompositions. Beyond universal constructions, the same synthesis engine adapts to structured inputs, automatically discovering structure-specific reductions independent of any prior assumptions about the matrix structure. For analytically structured matrices such as Householder reflectors, the system discovers a dimension-independent rule requiring only $2N-3$ MZIs, reducing complexity from quadratic to linear scaling, and generalizes to arbitrary $N$ without retraining. For matrices obtained from the SVD of sparse matrices, it uncovers structure-dependent reductions that tend to increase with sparsity, reaching up to $38\%$ fewer MZIs than the universal theoretical bound $N(N-1)/2$ at $95\%$ sparsity. Together, these results position program synthesis as a unified framework for both discovering generalizable decomposition rules and exploiting matrix-specific structure, enabling scalable automated design of photonic unitary circuits.
	
	As shown in Fig.~\ref{fig:main_results}\textbf{(a)}, our learning objective is to discover programs that transform arbitrary unitary matrices into diagonal form. This formulation directly corresponds to photonic implementation: synthesizing an arbitrary unitary matrix $U$ on a photonic chip decomposes into two steps---an MZI cascade network $P$ and a diagonal matrix $D$ such that $U = P^\dagger D$. Since $D$ requires no algorithmic search, our learning objective simplifies to finding the sequence $P$ that performs the diagonalization. Training tasks consist of randomly generated unitary matrices $U_k$~\cite{mezzadri2006generate}. A program $p$ transforms $U_k$ into a diagonal matrix $D_k = p(U_k)$, where $D_k$ retains phase information $e^{i\phi_j}$ on its diagonals. Success is determined solely by the diagonalization constraint: specifically, all off-diagonal elements ($i \neq j$) must satisfy $|p(U_k)_{ij}| < \epsilon_{\text{elem}}$, where $\epsilon_{\text{elem}} = 5 \times 10^{-4}$ in the simulation. The specific value of this threshold does not affect program acceptance, as successfully synthesized programs typically drive residuals down to machine precision ($\sim 10^{-16}$). Meanwhile, the diagonal phase values in $D_k$ are not prescribed, but emerge naturally from the decomposition process.

	To solve this diagonalization task, we define a set of $N(N-1)$ MZI primitives $\{R_{ij}, L_{ij}\}$ for $i > j$, with subscripts denoting the target element position at row $i$, column $j$. Each primitive eliminates one matrix element. The decomposition strategy exploits the unitarity constraint $U^\dagger U = I$, where systematic elimination of lower-triangular elements automatically enforces upper-triangular nullity, yielding a diagonal matrix. Each MZI primitive corresponds to a beam splitter transformation acting on adjacent modes, eliminating matrix element $u_{ij}$ via adaptive phase shifts. For $R_{ij}$-type primitives, the transformation eliminates element $u_{ij}$ by mixing columns $j$ and $j+1$ via right multiplication. From the nullification condition $(U R_{ij})_{ij} = 0$, the required phase $\theta_{ij}$ and mixing angle $\omega_{ij}$ are analytically derived: the ratio $r = u_{ij}/u_{i,j+1}$ is computed, with $\theta_{ij} = \arg(r)$ compensating the phase difference and $\omega_{ij} = \arctan(|r|)$ determining the mixing ratio. This constructs an $N \times N$ transformation matrix by embedding the $2 \times 2$ MZI block
	\begin{equation}
	M_{R,ij} = \begin{pmatrix} e^{-i\theta_{ij}}\cos\omega_{ij} & e^{-i\theta_{ij}}\sin\omega_{ij} \\ -\sin\omega_{ij} & \cos\omega_{ij} \end{pmatrix}
	\end{equation}
	at positions $(j, j+1)$ within an identity matrix, applying it via right multiplication $U' = U R_{ij}$. $L_{ij}$-type primitives similarly mix rows $i$ and $i-1$ via left multiplication, deriving parameters from condition $(L_{ij} U)_{ij} = 0$ and embedding the $2 \times 2$ MZI block
	\begin{equation}
	M_{L,ij} = \begin{pmatrix} e^{i\theta_{ij}}\cos\omega_{ij} & -\sin\omega_{ij} \\ e^{i\theta_{ij}}\sin\omega_{ij} & \cos\omega_{ij} \end{pmatrix}
	\end{equation}
	at positions $(i-1, i)$ within an identity matrix, applying it via left multiplication $U' = L_{ij} U$. The complete $N \times N$ matrix forms of $R_{ij}$ and $L_{ij}$ primitives are provided in Supplemental Material (SM). Each primitive takes a complex matrix as input and returns a transformed matrix, so primitives can be freely chained into sequences: e.g., a $3 \times 3$ decomposition might be \texttt{(lambda (R21 (L10 (L20 \$0))))} [applied in sequence to positions $(2,0)$, $(1,0)$, and $(2,1)$], where \texttt{\$0} denotes the input matrix and operations are applied from innermost to outermost.

	Given the task definition and primitive operations, we employ program synthesis to autonomously discover effective decomposition strategies. We adopt the DreamCoder framework~\cite{ellis2021dreamcoder}, which operates through an iterative wake-sleep cycle, as illustrated in Fig.~\ref{fig:main_results}\textbf{(b)}. During the wake phase, the system searches over candidate MZI operation sequences using the current primitive library and evaluates them against decomposition tasks. During the sleep phase, it identifies recurring patterns across successful solutions and abstracts them into new reusable primitives, expanding the library. The updated library is then carried into the next wake phase, where the richer primitive vocabulary enables more efficient search over deeper program structures~\cite{hinton1995wake}. We extend DreamCoder to the complex-valued matrix decomposition domain; implementation details are provided in SM. The MZI primitives $\{R_{ij}, L_{ij}\}$ encode physically meaningful beam-splitter transformations with analytically derived phase and rotation angles. This design can be generalized to a broad class of quantum information processing tasks involving matrix operations, such as quantum circuit synthesis~\cite{sarra2024discovering,ruiz2025quantum}.
	
	Verification of the synthesized programs depends on their target application. Universal decompositions are evaluated on multiple randomly generated unitary matrices, whereas structure-specific programs (e.g., for sparse matrices) are tested directly on their original inputs. In all cases, successfully verified programs achieve off-diagonal residuals at machine precision ($\sim 10^{-16}$). Furthermore, we validate the cross-dimensional generalization of universal programs by applying the learned patterns to larger matrices, confirming their ability to scale to arbitrary $N$.
	
\begin{figure}[h!]
		\centering
		\includegraphics[width=\columnwidth]{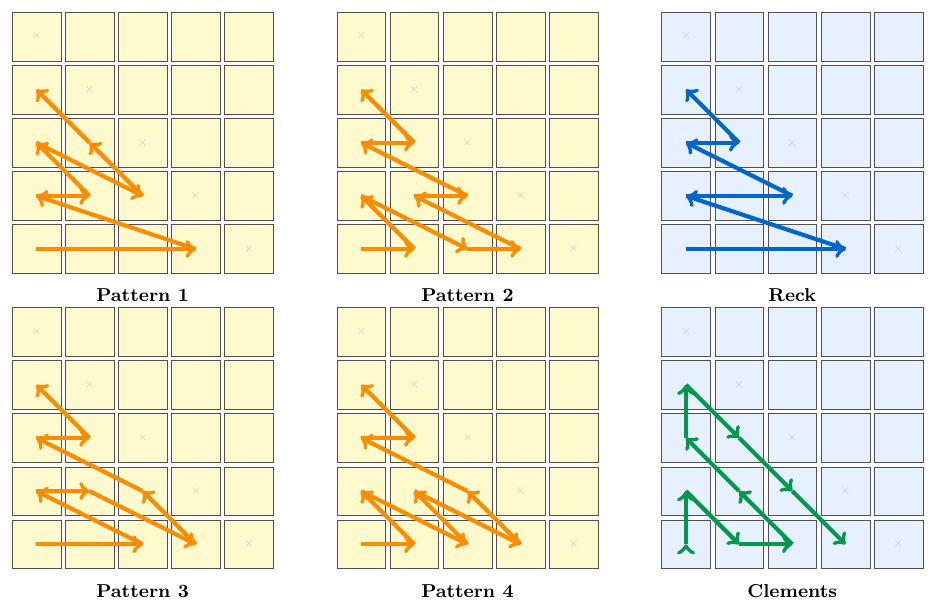}
		\caption{
			Decomposition patterns discovered for $5 \times 5$ unitary matrices compared with classical reference schemes. Left four panels (orange paths, yellow background) show representative discovered patterns exhibiting diverse elimination orderings. Right two panels show classical reference schemes for comparison: Reck (top, blue path) uses bottom-up row-wise elimination, while Clements (bottom, green path) employs symmetric alternating diagonal pattern. All patterns achieve the theoretical minimum of 10 MZI operations. The discovered patterns demonstrate that the system learned flexible decomposition strategies different from classical approaches.
		}
		\label{fig:n5patterns}
	\end{figure}

	Having established the framework, we now demonstrate its effectiveness on unitary decomposition tasks. To assess the system's capacity for autonomous algorithmic discovery, we employ independent learning where the system constructs complete decompositions from scratch without prior knowledge of classical schemes. This approach ensures discovered strategies emerge from the search process itself rather than reflecting biases from known methods. We first investigate the $5 \times 5$ case using independent learning within tractable search spaces. After nine wake-sleep iterations, the system autonomously discovered multiple distinct decomposition strategies, all achieving the theoretical minimum of 10 MZI operations. Figure~\ref{fig:n5patterns} visualizes four random decomposition patterns, where the orange paths trace the sequence of operations through the matrix. Each pattern shows a different execution ordering that systematically eliminates lower-triangular elements, confirming the system learned flexible decomposition strategies rather than a single fixed pattern. These strategies exhibit novel elimination orderings distinct from both Reck and Clements schemes, as presented in Fig.~\ref{fig:n5patterns}. The learned decomposition strategies are not tied to specific unitary instances but apply universally to arbitrary unitary matrices, functioning as generalizable algorithmic rules. To the best of our knowledge, such strategies have not been reported previously. Additional representative examples are provided in Table~S1 in SM.
	
	Among the discovered programs, the one ranked highest by the system, which favors solutions that are both correct and structurally simple (i.e., expressible more compactly in terms of the learned library operations)~\cite{lake2015human,liang2010learning} is \texttt{(lambda (R10 (R21 (R20 (R32 (R31 (R30 (R43 (R42 (R41 (L40 \$0)))))))))))}. This program corresponds to the matrix equation:
	\begin{equation}
	D = L_{40} \, U \, R_{41} R_{42} R_{43} R_{30} R_{31} R_{32} R_{20} R_{21} R_{10}\, ,
	\end{equation}
	where $D$ is the resulting diagonal matrix. Inverting this relation yields the unitary decomposition
	\begin{equation}
	U = L_{40}^{-1} D \, R_{10}^{-1} R_{21}^{-1} R_{20}^{-1} R_{32}^{-1} R_{31}^{-1} R_{30}^{-1} R_{43}^{-1} R_{42}^{-1} R_{41}^{-1}\, .
	\end{equation}
	This program directly maps to a photonic circuit implementation, where the five optical waveguides correspond to matrix modes and each MZI block executes one primitive operation. The circuit processes light through a cascade of 10 MZI stages, achieving the theoretical minimum count. This demonstrates that the system has learned a flexible decomposition strategy, revealing that the systematic elimination of lower-triangular elements admits multiple valid operation sequences.

	\begin{figure*}[t!]
		\centering
		\begin{minipage}[t]{0.48\textwidth}
			\vspace{0pt}
			\llap{\textbf{(a)}\hspace{0.3em}}%
			\includegraphics[width=\textwidth]{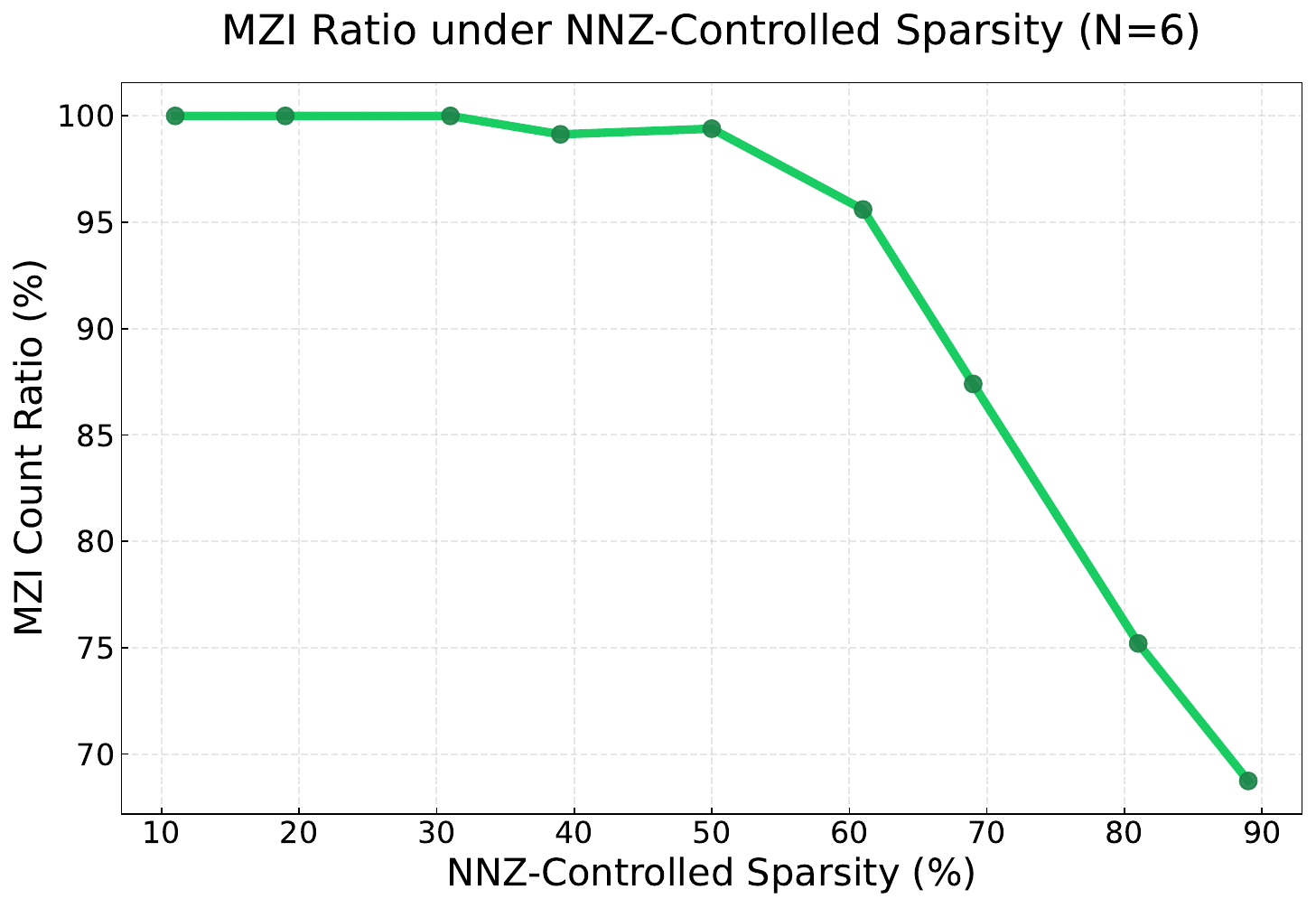}
		\end{minipage}
		\hfill
		\begin{minipage}[t]{0.48\textwidth}
			\vspace{0pt}
			\llap{\textbf{(b)}\hspace{0.3em}}%
			\includegraphics[width=\textwidth]{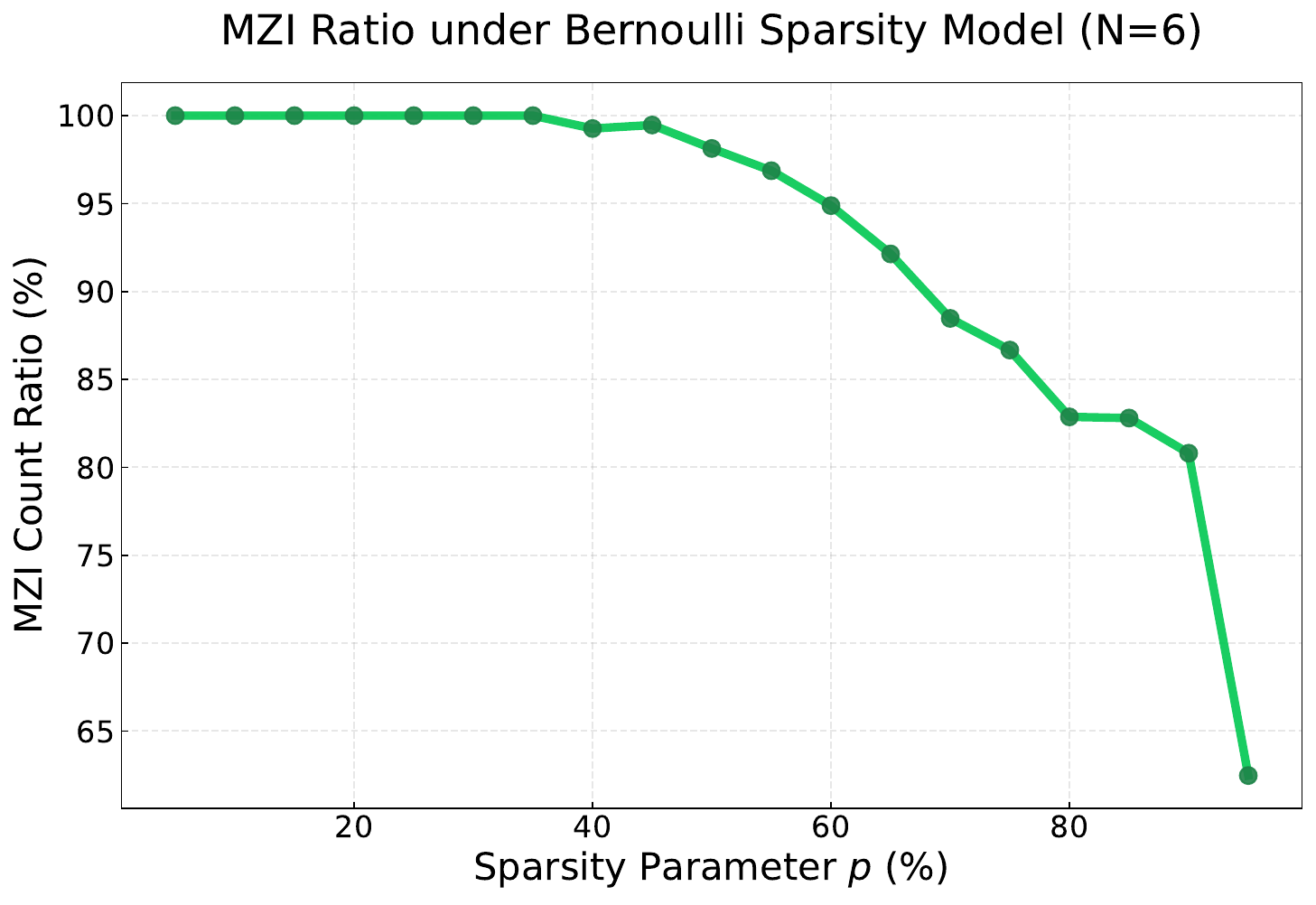}
		\end{minipage}
		\caption{MZI count ratio [normalized to the theoretical maximum $N(N-1)/2$] as a function of sparsity level for $N=6$ matrices under NNZ-controlled and Bernoulli sparsity models. Both models show a consistent overall decrease, confirming that the MZI reduction reflects an intrinsic property of sparse unitary matrices. Each data point is averaged over nearly 100 successfully decomposed matrices per sparsity level.}
		\label{fig:sparse_both}
	\end{figure*}
	
	Critically, the learned patterns exhibit cross-dimensional reasoning. Analysis of discovered programs reveals dimension-agnostic principles: systematic elimination strategies that remain valid across matrix scales. For instance, algorithmic abstractions derived from $5 \times 5$ matrices exhibit compositional generalization: they can be directly applied to higher-dimensional unitary matrices without retraining or modification. As an illustration, consider Pattern~4 shown in Fig.~\ref{fig:n5patterns}, which implements a row-pair interleaving strategy. Starting from the largest row index $i = N-1$ and decrementing by two each step, the algorithm processes rows in pairs: it first emits primitive $\mathtt{R}_{i,0}$, then interleaves primitives from two adjacent rows as $\mathtt{R}_{i,j} \to \mathtt{R}_{i-1,\,j-1}$ for $j = 1, \ldots, i-1$. When a single row remains ($i=1$), $\mathtt{R}_{1,0}$ closes the sequence. This fixed rule, discovered on $5 \times 5$ matrices, applies unchanged at any $N$ and enumerates all $N(N-1)/2$ operations exactly. Full pseudocode and generalization results are given in SM. We validated this cross-dimensional generalization by testing programs discovered on $5 \times 5$ training instances against progressively larger matrices, including $8 \times 8$, $16 \times 16$, $32 \times 32$, and $64 \times 64$ dimensions. All test cases achieve successful diagonalization at machine precision ($\sim 10^{-16}$), confirming that the learned decomposition strategy scales correctly to arbitrary dimensions. 	
	
	While our learned patterns exhibit dimension-agnostic principles that allow $5 \times 5$ strategies to generalize, direct training on $6 \times 6$ matrices remains essential to uncover richer decomposition orderings inaccessible from smaller examples. However, this objective confronts a fundamental bottleneck: the search space grows exponentially with matrix dimension, making the direct enumeration of deep 15-MZI sequences intractable. As matrix dimension grows, both the primitive vocabulary and program depth increase, causing the search space to explode combinatorially. To address this challenge, we employ curriculum learning~\cite{bengio2009curriculum,wang2021survey} by decomposing the problem into subtasks $\{T_3, T_2, T_1, T_0\}$. This progression allows the model to first master short subsequences in early stages ($T_3$), rapidly abstracting reusable patterns into library primitives via the sleep phase. Consequently, these consolidated abstractions effectively reduce the complexity for later stages ($T_0$), enabling the synthesis of deeply nested programs. Specifically, our results demonstrate that after just seven wake-sleep iterations, the system not only converged to the theoretical minimum depth but also expanded its library from 30 to 65 hierarchical abstractions. The system successfully rediscovered the Reck decomposition and identified novel patterns distinct from $5 \times 5$ independent learning. One such pattern can be expressed by the diagonalization relation \(D = L_{5,1}L_{4,0}L_{5,0}\,U\,\allowbreak R_{5,2}R_{5,3}R_{5,4}R_{4,1}R_{4,2}R_{3,0}R_{4,3}\allowbreak R_{3,1}R_{2,0}R_{3,2}R_{2,1}R_{1,0}\,\).
	This demonstrates that curriculum learning effectively navigates the expanded search space to uncover both classical and genuinely new decomposition strategies.

	Beyond universal decompositions, the system also discovers structure-aware reductions for specific matrix classes, without requiring any specialized analytical rules. We identify two broad categories of such structure-aware reductions.
	
	The first category comprises analytically structured matrices, with Householder matrices serving as a prime example. Defined as $H = I - 2vv^\dagger$ ($I$ is the identity matrix, $v \in \mathbb{C}^N$, $\|v\|=1$), Householder matrices are unitary reflectors widely used in mathematics and physics for stable unitary transformations, including tridiagonalization of Hamiltonians and eigenvalue calculations in quantum and many-body systems~\cite{golub2013matrix,ivanov2006engineering,lu2025experimental}. By exploiting these inherent structural constraints, our system autonomously identifies a dimension-independent decomposition rule that strictly bounds the total MZI count to $2N-3$. The recursive logic is as follows: extending to $N+1$ dimensions requires prepending two operations, $\mathtt{L}_{N,0}$ followed by $\mathtt{R}_{N,N-1}$, to the $N$-dimensional sequence. At $N=6$, the corresponding diagonalization relation is $D = L_{4,0}L_{5,0}\,H\,R_{5,4}R_{4,3}R_{3,0}R_{3,1}R_{3,2}R_{2,1}R_{1,0}$ (9 operations $= 2\times6-3$). This strategy reduces the hardware scaling from the universal quadratic bound of $N(N-1)/2$ to a linear complexity. When evaluated at $N=64$, this generalized rule perfectly reconstructs the unitary using only 125 MZIs instead of 2016. This represents a substantial 93.8\% reduction in hardware components, verified at machine precision without any retraining.
	
	The second category involves sparse matrices. In our framework, an arbitrary input $W$ is decomposed via SVD ($W = U\Sigma V^\dagger$), where the unitary factors $U$ and $V^\dagger$ inherit the sparsity of $W$ and are further decomposed into MZI sequences. Our system exploits this inherited structure to discover highly efficient MZI sequences. We validate this on $N=6$ matrices using two sparsity models: an NNZ-controlled setting (NNZ: number of nonzero elements) with a fixed number of nonzero elements in $W$, and a Bernoulli model where each element of $W$ is independently zeroed with probability $p$~\cite{markowitz1957elimination,gilbert1992sparse}. As shown in Fig.~\ref{fig:sparse_both}, the system achieves significant reductions relative to the universal theoretical bound $N(N-1)/2$, and these reductions generally increase with matrix sparsity. Under the NNZ-controlled model, reductions begin around $39\%$ sparsity and reach up to $31\%$ at $89\%$ sparsity. Under the Bernoulli model, reductions emerge around $40\%$ sparsity and grow more steeply at high sparsity, reaching up to $38\%$ at $95\%$ sparsity. This overall trend is consistent across both models, confirming that the efficiency gains reflect intrinsic structural properties of sparse unitaries rather than artifacts of specific matrix generation. For arbitrary unstructured sparse matrices, the discovered patterns are instance-specific, as there is no well-defined correspondence between sparsity patterns at different dimensions.
	
	These MZI reductions can provide practical hardware benefits: fewer components can help reduce optical insertion loss~\cite{clements2016optimal}, calibration complexity~\cite{lin2025high}, and power consumption~\cite{mojaver2023addressing}. Although synthesized topologies may be irregular, modern photonic design flows increasingly support automated placement and routing under physical constraints such as bend radius and waveguide crossings~\cite{zhou2025apollo,zhou2024automated}.

	In conclusion, our results show that program synthesis can uncover generalizable algorithmic structures for universal unitaries—going beyond known schemes to reveal decomposition principles and resource optimizations that classical Reck and Clements designs cannot capture. Built on a complex-valued matrix decomposition domain, our framework combines symbolic reasoning with physically grounded MZI primitives to systematically generate and verify decomposition programs, providing a scalable foundation that can be extended to broader quantum information tasks. These automatically generated constructions encode systematic construction methods, demonstrating the potential of AI-driven algorithm discovery to reveal fundamental principles and offering a powerful new approach for advancing photonic computing and quantum information processing.

	\textit{Acknowledgments}---This work is supported by the National Key Research and Development Program of China under Contract No. 2024YFA1610900 and the National Natural Science Foundation of China (NSFC) under Contract No. 12447106, No. 12541501, No. 12547102 and No. 12450404. D. D. H. also acknowledges the support of NSFC under No. 11875133 and No. 11075057, the National Key Research and Development Program of China under Contract No. 2018YFB2101302. The computations in this research were performed using the CFFF platform of Fudan University.

	\bibliographystyle{apsrev4-2-proof}
	\bibliography{references}%

\begin{thebibliography}{41}%
\makeatletter
\providecommand \@ifxundefined [1]{%
 \@ifx{#1\undefined}
}%
\providecommand \@ifnum [1]{%
 \ifnum #1\expandafter \@firstoftwo
 \else \expandafter \@secondoftwo
 \fi
}%
\providecommand \@ifx [1]{%
 \ifx #1\expandafter \@firstoftwo
 \else \expandafter \@secondoftwo
 \fi
}%
\providecommand \natexlab [1]{#1}%
\providecommand \enquote  [1]{``#1''}%
\providecommand \bibnamefont  [1]{#1}%
\providecommand \bibfnamefont [1]{#1}%
\providecommand \citenamefont [1]{#1}%
\providecommand \href@noop [0]{\@secondoftwo}%
\providecommand \href [0]{\begingroup \@sanitize@url \@href}%
\providecommand \@href[1]{\@@startlink{#1}\@@href}%
\providecommand \@@href[1]{\endgroup#1\@@endlink}%
\providecommand \@sanitize@url [0]{\catcode `\\12\catcode `\$12\catcode
  `\&12\catcode `\#12\catcode `\^12\catcode `\_12\catcode `\%12\relax}%
\providecommand \@@startlink[1]{}%
\providecommand \@@endlink[0]{}%
\providecommand \url  [0]{\begingroup\@sanitize@url \@url }%
\providecommand \@url [1]{\endgroup\@href {#1}{\urlprefix }}%
\providecommand \urlprefix  [0]{URL }%
\providecommand \Eprint [0]{\href }%
\providecommand \doibase [0]{https://doi.org/}%
\providecommand \selectlanguage [0]{\@gobble}%
\providecommand \bibinfo  [0]{\@secondoftwo}%
\providecommand \bibfield  [0]{\@secondoftwo}%
\providecommand \translation [1]{[#1]}%
\providecommand \BibitemOpen [0]{}%
\providecommand \bibitemStop [0]{}%
\providecommand \bibitemNoStop [0]{.\EOS\space}%
\providecommand \EOS [0]{\spacefactor3000\relax}%
\providecommand \BibitemShut  [1]{\csname bibitem#1\endcsname}%
\let\auto@bib@innerbib\@empty
\bibitem [{\citenamefont {Nielsen}\ and\ \citenamefont
  {Chuang}(2010)}]{nielsen2010quantum}%
  \BibitemOpen
  \bibfield  {author} {\bibinfo {author} {\bibfnamefont {M.~A.}\ \bibnamefont
  {Nielsen}}\ and\ \bibinfo {author} {\bibfnamefont {I.~L.}\ \bibnamefont
  {Chuang}},\ }\href@noop {} {\emph {\bibinfo {title} {{Quantum Computation and
  Quantum Information}}}}\ (\bibinfo  {publisher} {Cambridge University
  Press},\ \bibinfo {address} {Cambridge, England},\ \bibinfo {year}
  {2010})\BibitemShut {NoStop}%
\bibitem [{\citenamefont {Kok}\ \emph {et~al.}(2007)\citenamefont {Kok},
  \citenamefont {Munro}, \citenamefont {Nemoto}, \citenamefont {Ralph},
  \citenamefont {Dowling},\ and\ \citenamefont {Milburn}}]{kok2007linear}%
  \BibitemOpen
  \bibfield  {author} {\bibinfo {author} {\bibfnamefont {P.}~\bibnamefont
  {Kok}}, \bibinfo {author} {\bibfnamefont {W.~J.}\ \bibnamefont {Munro}},
  \bibinfo {author} {\bibfnamefont {K.}~\bibnamefont {Nemoto}}, \bibinfo
  {author} {\bibfnamefont {T.~C.}\ \bibnamefont {Ralph}}, \bibinfo {author}
  {\bibfnamefont {J.~P.}\ \bibnamefont {Dowling}},\ and\ \bibinfo {author}
  {\bibfnamefont {G.~J.}\ \bibnamefont {Milburn}},\ }\bibfield  {title}
  {\bibinfo {title} {Linear optical quantum computing with photonic qubits},\
  }\href@noop {} {\bibfield  {journal} {\bibinfo  {journal} {Rev. Mod. Phys.}\
  }\textbf {\bibinfo {volume} {79}},\ \bibinfo {pages} {135} (\bibinfo {year}
  {2007})}\BibitemShut {NoStop}%
\bibitem [{\citenamefont {Shastri}\ \emph {et~al.}(2021)\citenamefont
  {Shastri}, \citenamefont {Tait}, \citenamefont {Ferreira~de Lima},
  \citenamefont {Pernice}, \citenamefont {Bhaskaran}, \citenamefont {Wright},\
  and\ \citenamefont {Prucnal}}]{shastri2021photonics}%
  \BibitemOpen
  \bibfield  {author} {\bibinfo {author} {\bibfnamefont {B.~J.}\ \bibnamefont
  {Shastri}}, \bibinfo {author} {\bibfnamefont {A.~N.}\ \bibnamefont {Tait}},
  \bibinfo {author} {\bibfnamefont {T.}~\bibnamefont {Ferreira~de Lima}},
  \bibinfo {author} {\bibfnamefont {W.~H.}\ \bibnamefont {Pernice}}, \bibinfo
  {author} {\bibfnamefont {H.}~\bibnamefont {Bhaskaran}}, \bibinfo {author}
  {\bibfnamefont {C.~D.}\ \bibnamefont {Wright}},\ and\ \bibinfo {author}
  {\bibfnamefont {P.~R.}\ \bibnamefont {Prucnal}},\ }\bibfield  {title}
  {\bibinfo {title} {Photonics for artificial intelligence and neuromorphic
  computing},\ }\href@noop {} {\bibfield  {journal} {\bibinfo  {journal} {Nat.
  Photonics}\ }\textbf {\bibinfo {volume} {15}},\ \bibinfo {pages} {102}
  (\bibinfo {year} {2021})}\BibitemShut {NoStop}%
\bibitem [{\citenamefont {McMahon}(2023)}]{mcmahon2023physics}%
  \BibitemOpen
  \bibfield  {author} {\bibinfo {author} {\bibfnamefont {P.~L.}\ \bibnamefont
  {McMahon}},\ }\bibfield  {title} {\bibinfo {title} {The physics of optical
  computing},\ }\href@noop {} {\bibfield  {journal} {\bibinfo  {journal} {Nat.
  Rev. Phys.}\ }\textbf {\bibinfo {volume} {5}},\ \bibinfo {pages} {717}
  (\bibinfo {year} {2023})}\BibitemShut {NoStop}%
\bibitem [{\citenamefont {Knill}\ \emph {et~al.}(2001)\citenamefont {Knill},
  \citenamefont {Laflamme},\ and\ \citenamefont {Milburn}}]{knill2001scheme}%
  \BibitemOpen
  \bibfield  {author} {\bibinfo {author} {\bibfnamefont {E.}~\bibnamefont
  {Knill}}, \bibinfo {author} {\bibfnamefont {R.}~\bibnamefont {Laflamme}},\
  and\ \bibinfo {author} {\bibfnamefont {G.~J.}\ \bibnamefont {Milburn}},\
  }\bibfield  {title} {\bibinfo {title} {A scheme for efficient quantum
  computation with linear optics},\ }\href@noop {} {\bibfield  {journal}
  {\bibinfo  {journal} {Nature (London)}\ }\textbf {\bibinfo {volume} {409}},\
  \bibinfo {pages} {46} (\bibinfo {year} {2001})}\BibitemShut {NoStop}%
\bibitem [{\citenamefont {Shen}\ \emph {et~al.}(2017)\citenamefont {Shen},
  \citenamefont {Harris}, \citenamefont {Skirlo}, \citenamefont {Prabhu},
  \citenamefont {Baehr-Jones}, \citenamefont {Hochberg}, \citenamefont {Sun},
  \citenamefont {Zhao}, \citenamefont {Larochelle}, \citenamefont {Englund}
  \emph {et~al.}}]{shen2017deep}%
  \BibitemOpen
  \bibfield  {author} {\bibinfo {author} {\bibfnamefont {Y.}~\bibnamefont
  {Shen}}, \bibinfo {author} {\bibfnamefont {N.~C.}\ \bibnamefont {Harris}},
  \bibinfo {author} {\bibfnamefont {S.}~\bibnamefont {Skirlo}}, \bibinfo
  {author} {\bibfnamefont {M.}~\bibnamefont {Prabhu}}, \bibinfo {author}
  {\bibfnamefont {T.}~\bibnamefont {Baehr-Jones}}, \bibinfo {author}
  {\bibfnamefont {M.}~\bibnamefont {Hochberg}}, \bibinfo {author}
  {\bibfnamefont {X.}~\bibnamefont {Sun}}, \bibinfo {author} {\bibfnamefont
  {S.}~\bibnamefont {Zhao}}, \bibinfo {author} {\bibfnamefont {H.}~\bibnamefont
  {Larochelle}}, \bibinfo {author} {\bibfnamefont {D.}~\bibnamefont {Englund}}
  \emph {et~al.},\ }\bibfield  {title} {\bibinfo {title} {Deep learning with
  coherent nanophotonic circuits},\ }\href@noop {} {\bibfield  {journal}
  {\bibinfo  {journal} {Nat. Photonics}\ }\textbf {\bibinfo {volume} {11}},\
  \bibinfo {pages} {441} (\bibinfo {year} {2017})}\BibitemShut {NoStop}%
\bibitem [{\citenamefont {P{\'e}rez}\ \emph {et~al.}(2017)\citenamefont
  {P{\'e}rez}, \citenamefont {Gasulla}, \citenamefont {Crudgington},
  \citenamefont {Thomson}, \citenamefont {Khokhar}, \citenamefont {Li},
  \citenamefont {Cao}, \citenamefont {Mashanovich},\ and\ \citenamefont
  {Capmany}}]{perez2017multipurpose}%
  \BibitemOpen
  \bibfield  {author} {\bibinfo {author} {\bibfnamefont {D.}~\bibnamefont
  {P{\'e}rez}}, \bibinfo {author} {\bibfnamefont {I.}~\bibnamefont {Gasulla}},
  \bibinfo {author} {\bibfnamefont {L.}~\bibnamefont {Crudgington}}, \bibinfo
  {author} {\bibfnamefont {D.~J.}\ \bibnamefont {Thomson}}, \bibinfo {author}
  {\bibfnamefont {A.~Z.}\ \bibnamefont {Khokhar}}, \bibinfo {author}
  {\bibfnamefont {K.}~\bibnamefont {Li}}, \bibinfo {author} {\bibfnamefont
  {W.}~\bibnamefont {Cao}}, \bibinfo {author} {\bibfnamefont {G.~Z.}\
  \bibnamefont {Mashanovich}},\ and\ \bibinfo {author} {\bibfnamefont
  {J.}~\bibnamefont {Capmany}},\ }\bibfield  {title} {\bibinfo {title}
  {Multipurpose silicon photonics signal processor core},\ }\href@noop {}
  {\bibfield  {journal} {\bibinfo  {journal} {Nat. Commun.}\ }\textbf {\bibinfo
  {volume} {8}},\ \bibinfo {pages} {636} (\bibinfo {year} {2017})}\BibitemShut
  {NoStop}%
\bibitem [{\citenamefont {Lin}\ \emph {et~al.}(2018)\citenamefont {Lin},
  \citenamefont {Rivenson}, \citenamefont {Yardimci}, \citenamefont {Veli},
  \citenamefont {Luo}, \citenamefont {Jarrahi},\ and\ \citenamefont
  {Ozcan}}]{lin2018all}%
  \BibitemOpen
  \bibfield  {author} {\bibinfo {author} {\bibfnamefont {X.}~\bibnamefont
  {Lin}}, \bibinfo {author} {\bibfnamefont {Y.}~\bibnamefont {Rivenson}},
  \bibinfo {author} {\bibfnamefont {N.~T.}\ \bibnamefont {Yardimci}}, \bibinfo
  {author} {\bibfnamefont {M.}~\bibnamefont {Veli}}, \bibinfo {author}
  {\bibfnamefont {Y.}~\bibnamefont {Luo}}, \bibinfo {author} {\bibfnamefont
  {M.}~\bibnamefont {Jarrahi}},\ and\ \bibinfo {author} {\bibfnamefont
  {A.}~\bibnamefont {Ozcan}},\ }\bibfield  {title} {\bibinfo {title}
  {All-optical machine learning using diffractive deep neural networks},\
  }\href@noop {} {\bibfield  {journal} {\bibinfo  {journal} {Science}\ }\textbf
  {\bibinfo {volume} {361}},\ \bibinfo {pages} {1004} (\bibinfo {year}
  {2018})}\BibitemShut {NoStop}%
\bibitem [{\citenamefont {Pai}\ \emph {et~al.}(2023)\citenamefont {Pai},
  \citenamefont {Sun}, \citenamefont {Hughes}, \citenamefont {Park},
  \citenamefont {Bartlett}, \citenamefont {Williamson}, \citenamefont {Minkov},
  \citenamefont {Milanizadeh}, \citenamefont {Abebe}, \citenamefont
  {Morichetti} \emph {et~al.}}]{pai2023experimentally}%
  \BibitemOpen
  \bibfield  {author} {\bibinfo {author} {\bibfnamefont {S.}~\bibnamefont
  {Pai}}, \bibinfo {author} {\bibfnamefont {Z.}~\bibnamefont {Sun}}, \bibinfo
  {author} {\bibfnamefont {T.~W.}\ \bibnamefont {Hughes}}, \bibinfo {author}
  {\bibfnamefont {T.}~\bibnamefont {Park}}, \bibinfo {author} {\bibfnamefont
  {B.}~\bibnamefont {Bartlett}}, \bibinfo {author} {\bibfnamefont {I.~A.}\
  \bibnamefont {Williamson}}, \bibinfo {author} {\bibfnamefont
  {M.}~\bibnamefont {Minkov}}, \bibinfo {author} {\bibfnamefont
  {M.}~\bibnamefont {Milanizadeh}}, \bibinfo {author} {\bibfnamefont
  {N.}~\bibnamefont {Abebe}}, \bibinfo {author} {\bibfnamefont
  {F.}~\bibnamefont {Morichetti}} \emph {et~al.},\ }\bibfield  {title}
  {\bibinfo {title} {Experimentally realized in situ backpropagation for deep
  learning in photonic neural networks},\ }\href@noop {} {\bibfield  {journal}
  {\bibinfo  {journal} {Science}\ }\textbf {\bibinfo {volume} {380}},\ \bibinfo
  {pages} {398} (\bibinfo {year} {2023})}\BibitemShut {NoStop}%
\bibitem [{\citenamefont {Xue}\ \emph {et~al.}(2024)\citenamefont {Xue},
  \citenamefont {Zhou}, \citenamefont {Xu}, \citenamefont {Yu}, \citenamefont
  {Dai},\ and\ \citenamefont {Fang}}]{xue2024fully}%
  \BibitemOpen
  \bibfield  {author} {\bibinfo {author} {\bibfnamefont {Z.}~\bibnamefont
  {Xue}}, \bibinfo {author} {\bibfnamefont {T.}~\bibnamefont {Zhou}}, \bibinfo
  {author} {\bibfnamefont {Z.}~\bibnamefont {Xu}}, \bibinfo {author}
  {\bibfnamefont {S.}~\bibnamefont {Yu}}, \bibinfo {author} {\bibfnamefont
  {Q.}~\bibnamefont {Dai}},\ and\ \bibinfo {author} {\bibfnamefont
  {L.}~\bibnamefont {Fang}},\ }\bibfield  {title} {\bibinfo {title} {Fully
  forward mode training for optical neural networks},\ }\href@noop {}
  {\bibfield  {journal} {\bibinfo  {journal} {Nature (London)}\ }\textbf
  {\bibinfo {volume} {632}},\ \bibinfo {pages} {280} (\bibinfo {year}
  {2024})}\BibitemShut {NoStop}%
\bibitem [{\citenamefont {Fu}\ \emph {et~al.}(2024)\citenamefont {Fu},
  \citenamefont {Zhang}, \citenamefont {Sun}, \citenamefont {Huang},
  \citenamefont {Xu}, \citenamefont {Yang}, \citenamefont {Zhu},\ and\
  \citenamefont {Chen}}]{fu2024optical}%
  \BibitemOpen
  \bibfield  {author} {\bibinfo {author} {\bibfnamefont {T.}~\bibnamefont
  {Fu}}, \bibinfo {author} {\bibfnamefont {J.}~\bibnamefont {Zhang}}, \bibinfo
  {author} {\bibfnamefont {R.}~\bibnamefont {Sun}}, \bibinfo {author}
  {\bibfnamefont {Y.}~\bibnamefont {Huang}}, \bibinfo {author} {\bibfnamefont
  {W.}~\bibnamefont {Xu}}, \bibinfo {author} {\bibfnamefont {S.}~\bibnamefont
  {Yang}}, \bibinfo {author} {\bibfnamefont {Z.}~\bibnamefont {Zhu}},\ and\
  \bibinfo {author} {\bibfnamefont {H.}~\bibnamefont {Chen}},\ }\bibfield
  {title} {\bibinfo {title} {Optical neural networks: {Progress} and
  challenges},\ }\href@noop {} {\bibfield  {journal} {\bibinfo  {journal}
  {Light Sci. Appl.}\ }\textbf {\bibinfo {volume} {13}},\ \bibinfo {pages}
  {263} (\bibinfo {year} {2024})}\BibitemShut {NoStop}%
\bibitem [{\citenamefont {Bogaerts}\ \emph {et~al.}(2020)\citenamefont
  {Bogaerts}, \citenamefont {P{\'e}rez}, \citenamefont {Capmany}, \citenamefont
  {Miller}, \citenamefont {Poon}, \citenamefont {Englund}, \citenamefont
  {Morichetti},\ and\ \citenamefont {Melloni}}]{bogaerts2020programmable}%
  \BibitemOpen
  \bibfield  {author} {\bibinfo {author} {\bibfnamefont {W.}~\bibnamefont
  {Bogaerts}}, \bibinfo {author} {\bibfnamefont {D.}~\bibnamefont {P{\'e}rez}},
  \bibinfo {author} {\bibfnamefont {J.}~\bibnamefont {Capmany}}, \bibinfo
  {author} {\bibfnamefont {D.~A.}\ \bibnamefont {Miller}}, \bibinfo {author}
  {\bibfnamefont {J.}~\bibnamefont {Poon}}, \bibinfo {author} {\bibfnamefont
  {D.}~\bibnamefont {Englund}}, \bibinfo {author} {\bibfnamefont
  {F.}~\bibnamefont {Morichetti}},\ and\ \bibinfo {author} {\bibfnamefont
  {A.}~\bibnamefont {Melloni}},\ }\bibfield  {title} {\bibinfo {title}
  {Programmable photonic circuits},\ }\href@noop {} {\bibfield  {journal}
  {\bibinfo  {journal} {Nature (London)}\ }\textbf {\bibinfo {volume} {586}},\
  \bibinfo {pages} {207} (\bibinfo {year} {2020})}\BibitemShut {NoStop}%
\bibitem [{\citenamefont {O'brien}(2007)}]{o2007optical}%
  \BibitemOpen
  \bibfield  {author} {\bibinfo {author} {\bibfnamefont {J.~L.}\ \bibnamefont
  {O'brien}},\ }\bibfield  {title} {\bibinfo {title} {Optical quantum
  computing},\ }\href@noop {} {\bibfield  {journal} {\bibinfo  {journal}
  {Science}\ }\textbf {\bibinfo {volume} {318}},\ \bibinfo {pages} {1567}
  (\bibinfo {year} {2007})}\BibitemShut {NoStop}%
\bibitem [{\citenamefont {Konno}\ \emph {et~al.}(2024)\citenamefont {Konno},
  \citenamefont {Asavanant}, \citenamefont {Hanamura}, \citenamefont
  {Nagayoshi}, \citenamefont {Fukui}, \citenamefont {Sakaguchi}, \citenamefont
  {Ide}, \citenamefont {China}, \citenamefont {Yabuno}, \citenamefont {Miki}
  \emph {et~al.}}]{konno2024logical}%
  \BibitemOpen
  \bibfield  {author} {\bibinfo {author} {\bibfnamefont {S.}~\bibnamefont
  {Konno}}, \bibinfo {author} {\bibfnamefont {W.}~\bibnamefont {Asavanant}},
  \bibinfo {author} {\bibfnamefont {F.}~\bibnamefont {Hanamura}}, \bibinfo
  {author} {\bibfnamefont {H.}~\bibnamefont {Nagayoshi}}, \bibinfo {author}
  {\bibfnamefont {K.}~\bibnamefont {Fukui}}, \bibinfo {author} {\bibfnamefont
  {A.}~\bibnamefont {Sakaguchi}}, \bibinfo {author} {\bibfnamefont
  {R.}~\bibnamefont {Ide}}, \bibinfo {author} {\bibfnamefont {F.}~\bibnamefont
  {China}}, \bibinfo {author} {\bibfnamefont {M.}~\bibnamefont {Yabuno}},
  \bibinfo {author} {\bibfnamefont {S.}~\bibnamefont {Miki}} \emph {et~al.},\
  }\bibfield  {title} {\bibinfo {title} {Logical states for fault-tolerant
  quantum computation with propagating light},\ }\href@noop {} {\bibfield
  {journal} {\bibinfo  {journal} {Science}\ }\textbf {\bibinfo {volume}
  {383}},\ \bibinfo {pages} {289} (\bibinfo {year} {2024})}\BibitemShut
  {NoStop}%
\bibitem [{\citenamefont {{PsiQuantum
  team}}(2025)}]{psiquantum2025manufacturable}%
  \BibitemOpen
  \bibfield  {author} {\bibinfo {author} {\bibnamefont {{PsiQuantum team}}},\
  }\bibfield  {title} {\bibinfo {title} {A manufacturable platform for photonic
  quantum computing},\ }\href@noop {} {\bibfield  {journal} {\bibinfo
  {journal} {Nature (London)}\ }\textbf {\bibinfo {volume} {641}},\ \bibinfo
  {pages} {876} (\bibinfo {year} {2025})}\BibitemShut {NoStop}%
\bibitem [{\citenamefont {Pelucchi}\ \emph {et~al.}(2022)\citenamefont
  {Pelucchi}, \citenamefont {Fagas}, \citenamefont {Aharonovich}, \citenamefont
  {Englund}, \citenamefont {Figueroa}, \citenamefont {Gong}, \citenamefont
  {Hannes}, \citenamefont {Liu}, \citenamefont {Lu}, \citenamefont {Matsuda}
  \emph {et~al.}}]{pelucchi2022potential}%
  \BibitemOpen
  \bibfield  {author} {\bibinfo {author} {\bibfnamefont {E.}~\bibnamefont
  {Pelucchi}}, \bibinfo {author} {\bibfnamefont {G.}~\bibnamefont {Fagas}},
  \bibinfo {author} {\bibfnamefont {I.}~\bibnamefont {Aharonovich}}, \bibinfo
  {author} {\bibfnamefont {D.}~\bibnamefont {Englund}}, \bibinfo {author}
  {\bibfnamefont {E.}~\bibnamefont {Figueroa}}, \bibinfo {author}
  {\bibfnamefont {Q.}~\bibnamefont {Gong}}, \bibinfo {author} {\bibfnamefont
  {H.}~\bibnamefont {Hannes}}, \bibinfo {author} {\bibfnamefont
  {J.}~\bibnamefont {Liu}}, \bibinfo {author} {\bibfnamefont {C.-Y.}\
  \bibnamefont {Lu}}, \bibinfo {author} {\bibfnamefont {N.}~\bibnamefont
  {Matsuda}} \emph {et~al.},\ }\bibfield  {title} {\bibinfo {title} {The
  potential and global outlook of integrated photonics for quantum
  technologies},\ }\href@noop {} {\bibfield  {journal} {\bibinfo  {journal}
  {Nat. Rev. Phys.}\ }\textbf {\bibinfo {volume} {4}},\ \bibinfo {pages} {194}
  (\bibinfo {year} {2022})}\BibitemShut {NoStop}%
\bibitem [{\citenamefont {Bente}\ \emph {et~al.}(2025)\citenamefont {Bente},
  \citenamefont {Taheriniya}, \citenamefont {Lenzini}, \citenamefont
  {Br{\"u}ckerhoff-Pl{\"u}ckelmann}, \citenamefont {Kues}, \citenamefont
  {Bhaskaran}, \citenamefont {Wright},\ and\ \citenamefont
  {Pernice}}]{bente2025potential}%
  \BibitemOpen
  \bibfield  {author} {\bibinfo {author} {\bibfnamefont {I.}~\bibnamefont
  {Bente}}, \bibinfo {author} {\bibfnamefont {S.}~\bibnamefont {Taheriniya}},
  \bibinfo {author} {\bibfnamefont {F.}~\bibnamefont {Lenzini}}, \bibinfo
  {author} {\bibfnamefont {F.}~\bibnamefont {Br{\"u}ckerhoff-Pl{\"u}ckelmann}},
  \bibinfo {author} {\bibfnamefont {M.}~\bibnamefont {Kues}}, \bibinfo {author}
  {\bibfnamefont {H.}~\bibnamefont {Bhaskaran}}, \bibinfo {author}
  {\bibfnamefont {C.~D.}\ \bibnamefont {Wright}},\ and\ \bibinfo {author}
  {\bibfnamefont {W.}~\bibnamefont {Pernice}},\ }\bibfield  {title} {\bibinfo
  {title} {The potential of multidimensional photonic computing},\ }\href@noop
  {} {\bibfield  {journal} {\bibinfo  {journal} {Nat. Rev. Phys.}\ }\textbf
  {\bibinfo {volume} {7}},\ \bibinfo {pages} {439} (\bibinfo {year}
  {2025})}\BibitemShut {NoStop}%
\bibitem [{\citenamefont {Shekhar}\ \emph {et~al.}(2024)\citenamefont
  {Shekhar}, \citenamefont {Bogaerts}, \citenamefont {Chrostowski},
  \citenamefont {Bowers}, \citenamefont {Hochberg}, \citenamefont {Soref},\
  and\ \citenamefont {Shastri}}]{shekhar2024roadmapping}%
  \BibitemOpen
  \bibfield  {author} {\bibinfo {author} {\bibfnamefont {S.}~\bibnamefont
  {Shekhar}}, \bibinfo {author} {\bibfnamefont {W.}~\bibnamefont {Bogaerts}},
  \bibinfo {author} {\bibfnamefont {L.}~\bibnamefont {Chrostowski}}, \bibinfo
  {author} {\bibfnamefont {J.~E.}\ \bibnamefont {Bowers}}, \bibinfo {author}
  {\bibfnamefont {M.}~\bibnamefont {Hochberg}}, \bibinfo {author}
  {\bibfnamefont {R.}~\bibnamefont {Soref}},\ and\ \bibinfo {author}
  {\bibfnamefont {B.~J.}\ \bibnamefont {Shastri}},\ }\bibfield  {title}
  {\bibinfo {title} {Roadmapping the next generation of silicon photonics},\
  }\href@noop {} {\bibfield  {journal} {\bibinfo  {journal} {Nat. Commun.}\
  }\textbf {\bibinfo {volume} {15}},\ \bibinfo {pages} {751} (\bibinfo {year}
  {2024})}\BibitemShut {NoStop}%
\bibitem [{\citenamefont {Miller}(2013)}]{miller2013self}%
  \BibitemOpen
  \bibfield  {author} {\bibinfo {author} {\bibfnamefont {D.~A.}\ \bibnamefont
  {Miller}},\ }\bibfield  {title} {\bibinfo {title} {Self-configuring universal
  linear optical component},\ }\href@noop {} {\bibfield  {journal} {\bibinfo
  {journal} {Photonics Res.}\ }\textbf {\bibinfo {volume} {1}},\ \bibinfo
  {pages} {1} (\bibinfo {year} {2013})}\BibitemShut {NoStop}%
\bibitem [{\citenamefont {Carolan}\ \emph {et~al.}(2015)\citenamefont
  {Carolan}, \citenamefont {Harrold}, \citenamefont {Sparrow}, \citenamefont
  {Mart{\'\i}n-L{\'o}pez}, \citenamefont {Russell}, \citenamefont
  {Silverstone}, \citenamefont {Shadbolt}, \citenamefont {Matsuda},
  \citenamefont {Oguma}, \citenamefont {Itoh} \emph
  {et~al.}}]{carolan2015universal}%
  \BibitemOpen
  \bibfield  {author} {\bibinfo {author} {\bibfnamefont {J.}~\bibnamefont
  {Carolan}}, \bibinfo {author} {\bibfnamefont {C.}~\bibnamefont {Harrold}},
  \bibinfo {author} {\bibfnamefont {C.}~\bibnamefont {Sparrow}}, \bibinfo
  {author} {\bibfnamefont {E.}~\bibnamefont {Mart{\'\i}n-L{\'o}pez}}, \bibinfo
  {author} {\bibfnamefont {N.~J.}\ \bibnamefont {Russell}}, \bibinfo {author}
  {\bibfnamefont {J.~W.}\ \bibnamefont {Silverstone}}, \bibinfo {author}
  {\bibfnamefont {P.~J.}\ \bibnamefont {Shadbolt}}, \bibinfo {author}
  {\bibfnamefont {N.}~\bibnamefont {Matsuda}}, \bibinfo {author} {\bibfnamefont
  {M.}~\bibnamefont {Oguma}}, \bibinfo {author} {\bibfnamefont
  {M.}~\bibnamefont {Itoh}} \emph {et~al.},\ }\bibfield  {title} {\bibinfo
  {title} {Universal linear optics},\ }\href@noop {} {\bibfield  {journal}
  {\bibinfo  {journal} {Science}\ }\textbf {\bibinfo {volume} {349}},\ \bibinfo
  {pages} {711} (\bibinfo {year} {2015})}\BibitemShut {NoStop}%
\bibitem [{\citenamefont {Golub}\ and\ \citenamefont
  {Kahan}(1965)}]{golub1965calculating}%
  \BibitemOpen
  \bibfield  {author} {\bibinfo {author} {\bibfnamefont {G.}~\bibnamefont
  {Golub}}\ and\ \bibinfo {author} {\bibfnamefont {W.}~\bibnamefont {Kahan}},\
  }\bibfield  {title} {\bibinfo {title} {Calculating the singular values and
  pseudo-inverse of a matrix},\ }\href@noop {} {\bibfield  {journal} {\bibinfo
  {journal} {J. Soc. Ind. Appl. Math.}\ }\textbf {\bibinfo {volume} {2}},\
  \bibinfo {pages} {205} (\bibinfo {year} {1965})}\BibitemShut {NoStop}%
\bibitem [{\citenamefont {Reck}\ \emph {et~al.}(1994)\citenamefont {Reck},
  \citenamefont {Zeilinger}, \citenamefont {Bernstein},\ and\ \citenamefont
  {Bertani}}]{reck1994experimental}%
  \BibitemOpen
  \bibfield  {author} {\bibinfo {author} {\bibfnamefont {M.}~\bibnamefont
  {Reck}}, \bibinfo {author} {\bibfnamefont {A.}~\bibnamefont {Zeilinger}},
  \bibinfo {author} {\bibfnamefont {H.~J.}\ \bibnamefont {Bernstein}},\ and\
  \bibinfo {author} {\bibfnamefont {P.}~\bibnamefont {Bertani}},\ }\bibfield
  {title} {\bibinfo {title} {Experimental realization of any discrete unitary
  operator},\ }\href@noop {} {\bibfield  {journal} {\bibinfo  {journal} {Phys.
  Rev. Lett.}\ }\textbf {\bibinfo {volume} {73}},\ \bibinfo {pages} {58}
  (\bibinfo {year} {1994})}\BibitemShut {NoStop}%
\bibitem [{\citenamefont {Clements}\ \emph {et~al.}(2016)\citenamefont
  {Clements}, \citenamefont {Humphreys}, \citenamefont {Metcalf}, \citenamefont
  {Kolthammer},\ and\ \citenamefont {Walmsley}}]{clements2016optimal}%
  \BibitemOpen
  \bibfield  {author} {\bibinfo {author} {\bibfnamefont {W.~R.}\ \bibnamefont
  {Clements}}, \bibinfo {author} {\bibfnamefont {P.~C.}\ \bibnamefont
  {Humphreys}}, \bibinfo {author} {\bibfnamefont {B.~J.}\ \bibnamefont
  {Metcalf}}, \bibinfo {author} {\bibfnamefont {W.~S.}\ \bibnamefont
  {Kolthammer}},\ and\ \bibinfo {author} {\bibfnamefont {I.~A.}\ \bibnamefont
  {Walmsley}},\ }\bibfield  {title} {\bibinfo {title} {Optimal design for
  universal multiport interferometers},\ }\href@noop {} {\bibfield  {journal}
  {\bibinfo  {journal} {Optica}\ }\textbf {\bibinfo {volume} {3}},\ \bibinfo
  {pages} {1460} (\bibinfo {year} {2016})}\BibitemShut {NoStop}%
\bibitem [{\citenamefont {Ellis}\ \emph {et~al.}(2021)\citenamefont {Ellis},
  \citenamefont {Wong}, \citenamefont {Nye}, \citenamefont {Sabl{\'e}-Meyer},
  \citenamefont {Morales}, \citenamefont {Hewitt}, \citenamefont {Cary},
  \citenamefont {Solar-Lezama},\ and\ \citenamefont
  {Tenenbaum}}]{ellis2021dreamcoder}%
  \BibitemOpen
  \bibfield  {author} {\bibinfo {author} {\bibfnamefont {K.}~\bibnamefont
  {Ellis}}, \bibinfo {author} {\bibfnamefont {C.}~\bibnamefont {Wong}},
  \bibinfo {author} {\bibfnamefont {M.}~\bibnamefont {Nye}}, \bibinfo {author}
  {\bibfnamefont {M.}~\bibnamefont {Sabl{\'e}-Meyer}}, \bibinfo {author}
  {\bibfnamefont {L.}~\bibnamefont {Morales}}, \bibinfo {author} {\bibfnamefont
  {L.}~\bibnamefont {Hewitt}}, \bibinfo {author} {\bibfnamefont
  {L.}~\bibnamefont {Cary}}, \bibinfo {author} {\bibfnamefont {A.}~\bibnamefont
  {Solar-Lezama}},\ and\ \bibinfo {author} {\bibfnamefont {J.~B.}\ \bibnamefont
  {Tenenbaum}},\ }\bibfield  {title} {\bibinfo {title} {Dreamcoder:
  Bootstrapping inductive program synthesis with wake-sleep library learning},\
  }in\ \href@noop {} {\emph {\bibinfo {booktitle} {Proceedings of the 42nd {ACM
  SIGPLAN} International Conference on Programming Language Design and
  Implementation}}}\ (\bibinfo  {publisher} {Association for Computing
  Machinery (ACM)},\ \bibinfo {address} {New York},\ \bibinfo {year} {2021}),\
  pp.\ \bibinfo {pages} {835--850}\BibitemShut {NoStop}%
\bibitem [{\citenamefont {Mezzadri}(2007)}]{mezzadri2006generate}%
  \BibitemOpen
  \bibfield  {author} {\bibinfo {author} {\bibfnamefont {F.}~\bibnamefont
  {Mezzadri}},\ }\bibfield  {title} {\bibinfo {title} {How to generate random
  matrices from the classical compact groups},\ }\href@noop {} {\bibfield
  {journal} {\bibinfo  {journal} {Not. Am. Math. Soc.}\ }\textbf {\bibinfo
  {volume} {54}},\ \bibinfo {pages} {592} (\bibinfo {year} {2007})}\BibitemShut
  {NoStop}%
\bibitem [{\citenamefont {Hinton}\ \emph {et~al.}(1995)\citenamefont {Hinton},
  \citenamefont {Dayan}, \citenamefont {Frey},\ and\ \citenamefont
  {Neal}}]{hinton1995wake}%
  \BibitemOpen
  \bibfield  {author} {\bibinfo {author} {\bibfnamefont {G.~E.}\ \bibnamefont
  {Hinton}}, \bibinfo {author} {\bibfnamefont {P.}~\bibnamefont {Dayan}},
  \bibinfo {author} {\bibfnamefont {B.~J.}\ \bibnamefont {Frey}},\ and\
  \bibinfo {author} {\bibfnamefont {R.~M.}\ \bibnamefont {Neal}},\ }\bibfield
  {title} {\bibinfo {title} {The ``wake-sleep'' algorithm for unsupervised
  neural networks},\ }\href@noop {} {\bibfield  {journal} {\bibinfo  {journal}
  {Science}\ }\textbf {\bibinfo {volume} {268}},\ \bibinfo {pages} {1158}
  (\bibinfo {year} {1995})}\BibitemShut {NoStop}%
\bibitem [{\citenamefont {Sarra}\ \emph {et~al.}(2024)\citenamefont {Sarra},
  \citenamefont {Ellis},\ and\ \citenamefont
  {Marquardt}}]{sarra2024discovering}%
  \BibitemOpen
  \bibfield  {author} {\bibinfo {author} {\bibfnamefont {L.}~\bibnamefont
  {Sarra}}, \bibinfo {author} {\bibfnamefont {K.}~\bibnamefont {Ellis}},\ and\
  \bibinfo {author} {\bibfnamefont {F.}~\bibnamefont {Marquardt}},\ }\bibfield
  {title} {\bibinfo {title} {Discovering quantum circuit components with
  program synthesis},\ }\href@noop {} {\bibfield  {journal} {\bibinfo
  {journal} {Mach. Learn. Sci. Technol.}\ }\textbf {\bibinfo {volume} {5}},\
  \bibinfo {pages} {025029} (\bibinfo {year} {2024})}\BibitemShut {NoStop}%
\bibitem [{\citenamefont {Ruiz}\ \emph {et~al.}(2025)\citenamefont {Ruiz},
  \citenamefont {Laakkonen}, \citenamefont {Bausch}, \citenamefont {Balog},
  \citenamefont {Barekatain}, \citenamefont {Heras}, \citenamefont {Novikov},
  \citenamefont {Fitzpatrick}, \citenamefont {Romera-Paredes}, \citenamefont
  {Van De~Wetering} \emph {et~al.}}]{ruiz2025quantum}%
  \BibitemOpen
  \bibfield  {author} {\bibinfo {author} {\bibfnamefont {F.~J.}\ \bibnamefont
  {Ruiz}}, \bibinfo {author} {\bibfnamefont {T.}~\bibnamefont {Laakkonen}},
  \bibinfo {author} {\bibfnamefont {J.}~\bibnamefont {Bausch}}, \bibinfo
  {author} {\bibfnamefont {M.}~\bibnamefont {Balog}}, \bibinfo {author}
  {\bibfnamefont {M.}~\bibnamefont {Barekatain}}, \bibinfo {author}
  {\bibfnamefont {F.~J.}\ \bibnamefont {Heras}}, \bibinfo {author}
  {\bibfnamefont {A.}~\bibnamefont {Novikov}}, \bibinfo {author} {\bibfnamefont
  {N.}~\bibnamefont {Fitzpatrick}}, \bibinfo {author} {\bibfnamefont
  {B.}~\bibnamefont {Romera-Paredes}}, \bibinfo {author} {\bibfnamefont
  {J.}~\bibnamefont {Van De~Wetering}} \emph {et~al.},\ }\bibfield  {title}
  {\bibinfo {title} {Quantum circuit optimization with alphatensor},\
  }\href@noop {} {\bibfield  {journal} {\bibinfo  {journal} {Nat. Mach.
  Intell.}\ }\textbf {\bibinfo {volume} {7}},\ \bibinfo {pages} {374} (\bibinfo
  {year} {2025})}\BibitemShut {NoStop}%
\bibitem [{\citenamefont {Lake}\ \emph {et~al.}(2015)\citenamefont {Lake},
  \citenamefont {Salakhutdinov},\ and\ \citenamefont
  {Tenenbaum}}]{lake2015human}%
  \BibitemOpen
  \bibfield  {author} {\bibinfo {author} {\bibfnamefont {B.~M.}\ \bibnamefont
  {Lake}}, \bibinfo {author} {\bibfnamefont {R.}~\bibnamefont
  {Salakhutdinov}},\ and\ \bibinfo {author} {\bibfnamefont {J.~B.}\
  \bibnamefont {Tenenbaum}},\ }\bibfield  {title} {\bibinfo {title}
  {Human-level concept learning through probabilistic program induction},\
  }\href@noop {} {\bibfield  {journal} {\bibinfo  {journal} {Science}\ }\textbf
  {\bibinfo {volume} {350}},\ \bibinfo {pages} {1332} (\bibinfo {year}
  {2015})}\BibitemShut {NoStop}%
\bibitem [{\citenamefont {Liang}\ \emph {et~al.}(2010)\citenamefont {Liang},
  \citenamefont {Jordan},\ and\ \citenamefont {Klein}}]{liang2010learning}%
  \BibitemOpen
  \bibfield  {author} {\bibinfo {author} {\bibfnamefont {P.}~\bibnamefont
  {Liang}}, \bibinfo {author} {\bibfnamefont {M.~I.}\ \bibnamefont {Jordan}},\
  and\ \bibinfo {author} {\bibfnamefont {D.}~\bibnamefont {Klein}},\ }\bibfield
   {title} {\bibinfo {title} {Learning programs: A hierarchical {Bayesian}
  approach},\ }in\ \href@noop {} {\emph {\bibinfo {booktitle} {ICML}}}\
  (\bibinfo  {publisher} {Omnipress},\ \bibinfo {address} {Madison},\ \bibinfo
  {year} {2010}),\ Vol.~\bibinfo {volume} {10},\ pp.\ \bibinfo {pages}
  {639--646}\BibitemShut {NoStop}%
\bibitem [{\citenamefont {Bengio}\ \emph {et~al.}(2009)\citenamefont {Bengio},
  \citenamefont {Louradour}, \citenamefont {Collobert},\ and\ \citenamefont
  {Weston}}]{bengio2009curriculum}%
  \BibitemOpen
  \bibfield  {author} {\bibinfo {author} {\bibfnamefont {Y.}~\bibnamefont
  {Bengio}}, \bibinfo {author} {\bibfnamefont {J.}~\bibnamefont {Louradour}},
  \bibinfo {author} {\bibfnamefont {R.}~\bibnamefont {Collobert}},\ and\
  \bibinfo {author} {\bibfnamefont {J.}~\bibnamefont {Weston}},\ }\bibfield
  {title} {\bibinfo {title} {Curriculum learning},\ }in\ \href@noop {} {\emph
  {\bibinfo {booktitle} {Proceedings of the 26th annual international
  conference on machine learning}}}\ (\bibinfo  {publisher} {Omnipress},\
  \bibinfo {address} {Madison},\ \bibinfo {year} {2009}),\ pp.\ \bibinfo
  {pages} {41--48}\BibitemShut {NoStop}%
\bibitem [{\citenamefont {Wang}\ \emph {et~al.}(2021)\citenamefont {Wang},
  \citenamefont {Chen},\ and\ \citenamefont {Zhu}}]{wang2021survey}%
  \BibitemOpen
  \bibfield  {author} {\bibinfo {author} {\bibfnamefont {X.}~\bibnamefont
  {Wang}}, \bibinfo {author} {\bibfnamefont {Y.}~\bibnamefont {Chen}},\ and\
  \bibinfo {author} {\bibfnamefont {W.}~\bibnamefont {Zhu}},\ }\bibfield
  {title} {\bibinfo {title} {A survey on curriculum learning},\ }\href@noop {}
  {\bibfield  {journal} {\bibinfo  {journal} {IEEE Trans. Pattern Anal. Mach.
  Intell.}\ }\textbf {\bibinfo {volume} {44}},\ \bibinfo {pages} {4555}
  (\bibinfo {year} {2021})}\BibitemShut {NoStop}%
\bibitem [{\citenamefont {Golub}\ and\ \citenamefont
  {Van~Loan}(2013)}]{golub2013matrix}%
  \BibitemOpen
  \bibfield  {author} {\bibinfo {author} {\bibfnamefont {G.~H.}\ \bibnamefont
  {Golub}}\ and\ \bibinfo {author} {\bibfnamefont {C.~F.}\ \bibnamefont
  {Van~Loan}},\ }\href@noop {} {\emph {\bibinfo {title} {{Matrix
  Computations}}}}\ (\bibinfo  {publisher} {JHU Press},\ \bibinfo {address}
  {Baltimore, Maryland},\ \bibinfo {year} {2013})\BibitemShut {NoStop}%
\bibitem [{\citenamefont {Ivanov}\ \emph {et~al.}(2006)\citenamefont {Ivanov},
  \citenamefont {Kyoseva},\ and\ \citenamefont
  {Vitanov}}]{ivanov2006engineering}%
  \BibitemOpen
  \bibfield  {author} {\bibinfo {author} {\bibfnamefont {P.~A.}\ \bibnamefont
  {Ivanov}}, \bibinfo {author} {\bibfnamefont {E.}~\bibnamefont {Kyoseva}},\
  and\ \bibinfo {author} {\bibfnamefont {N.~V.}\ \bibnamefont {Vitanov}},\
  }\bibfield  {title} {\bibinfo {title} {Engineering of arbitrary {U (N)}
  transformations by quantum householder reflections},\ }\href@noop {}
  {\bibfield  {journal} {\bibinfo  {journal} {Phys. Rev. A}\ }\textbf {\bibinfo
  {volume} {74}},\ \bibinfo {pages} {022323} (\bibinfo {year}
  {2006})}\BibitemShut {NoStop}%
\bibitem [{\citenamefont {Lu}\ \emph {et~al.}(2025)\citenamefont {Lu},
  \citenamefont {Wang},\ and\ \citenamefont {Ma}}]{lu2025experimental}%
  \BibitemOpen
  \bibfield  {author} {\bibinfo {author} {\bibfnamefont {C.}~\bibnamefont
  {Lu}}, \bibinfo {author} {\bibfnamefont {X.}~\bibnamefont {Wang}},\ and\
  \bibinfo {author} {\bibfnamefont {G.}~\bibnamefont {Ma}},\ }\bibfield
  {title} {\bibinfo {title} {Experimental realization of special-unitary
  operations in classical mechanics by nonadiabatic evolutions},\ }\href@noop
  {} {\bibfield  {journal} {\bibinfo  {journal} {Phys. Rev. Lett.}\ }\textbf
  {\bibinfo {volume} {135}},\ \bibinfo {pages} {027201} (\bibinfo {year}
  {2025})}\BibitemShut {NoStop}%
\bibitem [{\citenamefont {Markowitz}(1957)}]{markowitz1957elimination}%
  \BibitemOpen
  \bibfield  {author} {\bibinfo {author} {\bibfnamefont {H.~M.}\ \bibnamefont
  {Markowitz}},\ }\bibfield  {title} {\bibinfo {title} {The elimination form of
  the inverse and its application to linear programming},\ }\href@noop {}
  {\bibfield  {journal} {\bibinfo  {journal} {Manage. Sci.}\ }\textbf {\bibinfo
  {volume} {3}},\ \bibinfo {pages} {255} (\bibinfo {year} {1957})}\BibitemShut
  {NoStop}%
\bibitem [{\citenamefont {Gilbert}\ \emph {et~al.}(1992)\citenamefont
  {Gilbert}, \citenamefont {Moler},\ and\ \citenamefont
  {Schreiber}}]{gilbert1992sparse}%
  \BibitemOpen
  \bibfield  {author} {\bibinfo {author} {\bibfnamefont {J.~R.}\ \bibnamefont
  {Gilbert}}, \bibinfo {author} {\bibfnamefont {C.}~\bibnamefont {Moler}},\
  and\ \bibinfo {author} {\bibfnamefont {R.}~\bibnamefont {Schreiber}},\
  }\bibfield  {title} {\bibinfo {title} {Sparse matrices in matlab: Design and
  implementation},\ }\href@noop {} {\bibfield  {journal} {\bibinfo  {journal}
  {SIAM J. Matrix Anal. Appl.}\ }\textbf {\bibinfo {volume} {13}},\ \bibinfo
  {pages} {333} (\bibinfo {year} {1992})}\BibitemShut {NoStop}%
\bibitem [{\citenamefont {Lin}\ \emph {et~al.}(2025)\citenamefont {Lin},
  \citenamefont {Zeng}, \citenamefont {Lin}, \citenamefont {Yu},\ and\
  \citenamefont {Zhang}}]{lin2025high}%
  \BibitemOpen
  \bibfield  {author} {\bibinfo {author} {\bibfnamefont {S.}~\bibnamefont
  {Lin}}, \bibinfo {author} {\bibfnamefont {J.}~\bibnamefont {Zeng}}, \bibinfo
  {author} {\bibfnamefont {S.}~\bibnamefont {Lin}}, \bibinfo {author}
  {\bibfnamefont {S.}~\bibnamefont {Yu}},\ and\ \bibinfo {author}
  {\bibfnamefont {Y.}~\bibnamefont {Zhang}},\ }\bibfield  {title} {\bibinfo
  {title} {High-fidelity and compact topology architecture for large-scale
  reconfigurable linear optical networks},\ }\href@noop {} {\bibfield
  {journal} {\bibinfo  {journal} {Adv. Photonics Nexus}\ }\textbf {\bibinfo
  {volume} {4}},\ \bibinfo {pages} {066012} (\bibinfo {year}
  {2025})}\BibitemShut {NoStop}%
\bibitem [{\citenamefont {Mojaver}\ \emph {et~al.}(2023)\citenamefont
  {Mojaver}, \citenamefont {Zhao}, \citenamefont {Leung}, \citenamefont
  {Safaee},\ and\ \citenamefont {Liboiron-Ladouceur}}]{mojaver2023addressing}%
  \BibitemOpen
  \bibfield  {author} {\bibinfo {author} {\bibfnamefont {K.~H.~R.}\
  \bibnamefont {Mojaver}}, \bibinfo {author} {\bibfnamefont {B.}~\bibnamefont
  {Zhao}}, \bibinfo {author} {\bibfnamefont {E.}~\bibnamefont {Leung}},
  \bibinfo {author} {\bibfnamefont {S.~M.~R.}\ \bibnamefont {Safaee}},\ and\
  \bibinfo {author} {\bibfnamefont {O.}~\bibnamefont {Liboiron-Ladouceur}},\
  }\bibfield  {title} {\bibinfo {title} {Addressing the programming challenges
  of practical interferometric mesh based optical processors},\ }\href@noop {}
  {\bibfield  {journal} {\bibinfo  {journal} {Opt. Express}\ }\textbf {\bibinfo
  {volume} {31}},\ \bibinfo {pages} {23851} (\bibinfo {year}
  {2023})}\BibitemShut {NoStop}%
\bibitem [{\citenamefont {Zhou}\ \emph {et~al.}(2025)\citenamefont {Zhou},
  \citenamefont {Yang}, \citenamefont {Gangi}, \citenamefont {Huang},
  \citenamefont {Ren},\ and\ \citenamefont {Gu}}]{zhou2025apollo}%
  \BibitemOpen
  \bibfield  {author} {\bibinfo {author} {\bibfnamefont {H.}~\bibnamefont
  {Zhou}}, \bibinfo {author} {\bibfnamefont {H.}~\bibnamefont {Yang}}, \bibinfo
  {author} {\bibfnamefont {N.}~\bibnamefont {Gangi}}, \bibinfo {author}
  {\bibfnamefont {Z.~R.}\ \bibnamefont {Huang}}, \bibinfo {author}
  {\bibfnamefont {H.}~\bibnamefont {Ren}},\ and\ \bibinfo {author}
  {\bibfnamefont {J.}~\bibnamefont {Gu}},\ }\bibfield  {title} {\bibinfo
  {title} {Apollo: Automated routing-informed placement for large-scale
  photonic integrated circuits},\ }in\ \href@noop {} {\emph {\bibinfo
  {booktitle} {2025 IEEE/ACM International Conference On Computer Aided Design
  (ICCAD)}}}\ (\bibinfo {organization} {IEEE},\ \bibinfo {address} {New York},\
  \bibinfo {year} {2025}),\ pp.\ \bibinfo {pages} {1--9}\BibitemShut {NoStop}%
\bibitem [{\citenamefont {Zhou}\ \emph {et~al.}()\citenamefont {Zhou},
  \citenamefont {Zhu},\ and\ \citenamefont {Gu}}]{zhou2024automated}%
  \BibitemOpen
  \bibfield  {author} {\bibinfo {author} {\bibfnamefont {H.}~\bibnamefont
  {Zhou}}, \bibinfo {author} {\bibfnamefont {K.}~\bibnamefont {Zhu}},\ and\
  \bibinfo {author} {\bibfnamefont {J.}~\bibnamefont {Gu}},\ }\href@noop {}
  {\bibinfo {title} {Automated curvy waveguide routing for large-scale photonic
  integrated circuits}},\ \bibinfo {note}
  {\href{https://arxiv.org/abs/2410.01260}{arXiv: 2410.01260}}\BibitemShut
  {NoStop}%
\end{thebibliography}%
	
	\clearpage
	\onecolumngrid

	\setcounter{page}{1}
	\setcounter{figure}{0}
	\setcounter{table}{0}
	\setcounter{equation}{0}
	\setcounter{section}{0}
	
	\renewcommand{\theequation}{S.\arabic{equation}} 
	
	\renewcommand{\thefigure}{S\arabic{figure}}
	
	\renewcommand{\thetable}{S\arabic{table}}

	\begin{center}
		\section*{Supplemental Material for ``Program-Synthesis–Driven Auto-Design of Universal Unitary Operators''}
	\end{center}
	
	\vspace{0.5cm}
	
	In this Supplemental Material, we present more details on the framework implementation, curriculum learning strategy, and synthesized programs.
	
	\vspace{0.5cm}
	
	\subsection*{FRAMEWORK AND IMPLEMENTATION DETAILS}
	
	This section provides a detailed exposition of the DreamCoder program synthesis framework~\cite{ellis2021dreamcoder} and our matrix decomposition domain implementation.
	
	The DreamCoder framework operates through alternating wake and sleep phases. During the wake phase, the system performs enumerative search over the space of programs expressible in the current grammar. Programs are enumerated in order of decreasing prior probability under a learned probabilistic context-free grammar (PCFG). This grammar assigns higher probabilities to operation sequences that have proven useful in previous iterations, guiding the search toward efficient decompositions. Programs are then evaluated against training tasks using a posterior-based scoring function. For each task, the search identifies programs that maximize the posterior probability $P(p|T) \propto P(T|p)P(p)$, where $P(T|p)$ measures how well program $p$ solves task $T$ and $P(p)$ is the prior probability assigned by the grammar (favoring programs with shorter description length under the current library). The wake phase produces a corpus of successful programs paired with the tasks they solve.
	
	During the sleep phase, the system analyzes the corpus of programs discovered during wake to identify frequently recurring substructures. It employs a compression-based abstraction algorithm that searches for program fragments whose factorization into reusable library functions would maximize the overall description length reduction across the entire corpus. Given a set of programs $\{p_1, ..., p_n\}$, the algorithm identifies a new abstraction $f$ and rewritten programs $\{p'_1, ..., p'_n\}$ that minimize the total cost $\text{cost}(f) + \sum_i \text{cost}(p'_i)$, where cost is measured by program size under the grammar. Successfully abstracted fragments are added to the library as new primitives, enabling the next wake phase to build upon these higher-level building blocks. The updated library is then used in the next wake phase, closing the learning loop and enabling progressively more efficient search over successive iterations.
	
	Program evaluation with hard constraints. For matrix decomposition tasks, we extend DreamCoder's evaluation mechanism from exact matching to constraint-based validation. Each task consists of unitary input matrices $\{U_k\}$, with success defined by transforming each input into diagonal form. A program $p$ succeeds when all off-diagonal elements satisfy $|p(U_k)_{ij}| < \epsilon_{\text{elem}} = 5 \times 10^{-4}$ for all $k$ and all $i \neq j$. In practice, successfully synthesized programs drive all off-diagonal elements to near machine precision ($\sim 10^{-16}$), so the specific value of $\epsilon_{\text{elem}}$ does not affect which programs are accepted. Programs passing this check are then ranked by posterior probability $P(p|T) \propto P(T|p)P(p)$, where the likelihood $P(T|p)$ is measured by the average off-diagonal Frobenius norm $\frac{1}{K}\sum_k \|p(U_k) - \mathrm{diag}(p(U_k))\|_F$, serving as a continuous score reflecting decomposition quality, and $P(p)$ is the prior from the learned grammar favoring shorter programs. In the tables of discovered programs, we report the log-posterior $\log P(p|T)$: a less negative value indicates a program that is both more accurate and has a shorter description length under the learned library grammar. The program with the shortest description length under the learned library that correctly solves all tasks receives the highest (least negative) log-posterior and is selected as the representative solution.
	
	The matrix decomposition domain extends DreamCoder through a two-component software interface. The Python frontend defines the MZI primitives, generates training tasks, and coordinates the learning loop. The OCaml backend performs the combinatorial search over program sequences, leveraging its efficient multicore enumeration capability. The two components communicate by passing matrix data and candidate programs as structured messages in JSON format (a lightweight, human-readable data interchange format). The Python component defines matrix decomposition tasks, constructs a parameterizable primitive library for $N \times N$ matrices, and supports curriculum-style task generation, while the OCaml backend implements the corresponding primitive semantics for matrix evaluation and verification and carries out program search based on these task and primitive specifications.
	
	\begin{figure}[h!]
		\centering
		\includegraphics[width=0.90\textwidth]{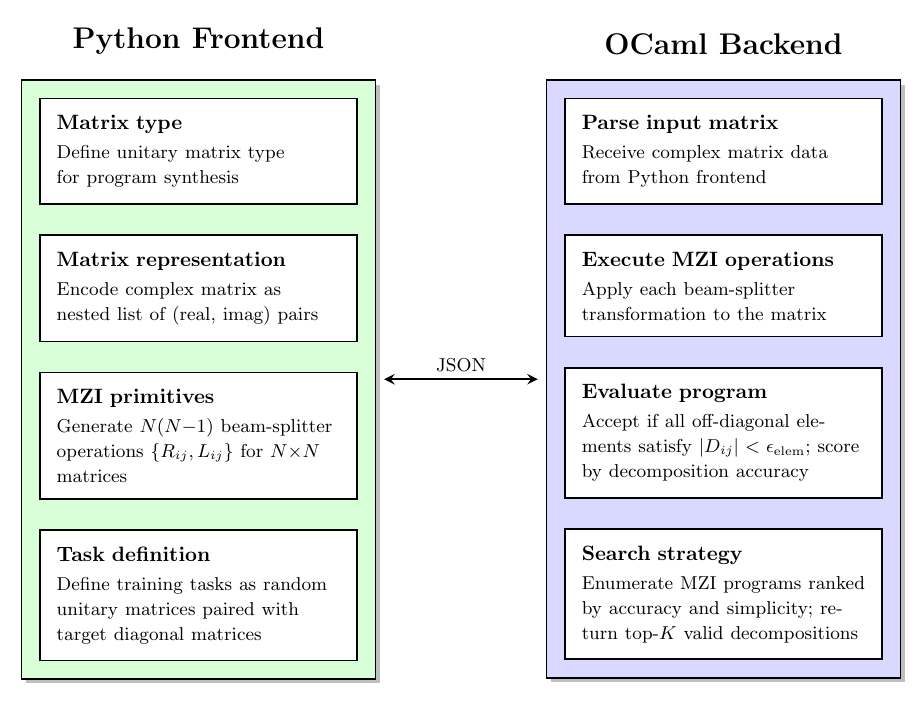}
		\caption{
			Program synthesis framework for unitary matrix decomposition. The system learns from unitary matrices $U$ and seeks to transform them into diagonal form, discovering MZI operation sequences via wake-sleep cycles. During waking, candidate MZI sequences are enumerated and evaluated against the diagonalization criterion. During sleep, recurring sub-sequences are abstracted into reusable building blocks and added to a growing library, enabling discovery of increasingly complex decompositions in later cycles.
		}
		\label{fig:framework_architecture}
	\end{figure}
	
	For matrix element elimination, two primitive families are defined: $R_{ij}$ and $L_{ij}$ for $i > j$. The $R_{ij}$ primitive targets element $u_{ij}$ via right multiplication (column operations), while $L_{ij}$ uses left multiplication (row operations). Both are parameterized by row and column indices through partial application.
	
	The $R_{ij}$ primitive constructs its transformation from element $u_{ij}$ and its right neighbor $u_{i,j+1}$, computing the complex ratio, phase angle, and rotation angle:
	\begin{align*}
	r &= u_{ij}/u_{i,j+1}, \quad \theta_{ij} = \arg(r), \quad \omega_{ij} = \arctan(|r|).
	\end{align*}
	The $2 \times 2$ MZI transformation matrix is:
	\begin{equation}
	M_{R,ij} = \begin{bmatrix} e^{-i\theta_{ij}}\cos\omega_{ij} & e^{-i\theta_{ij}}\sin\omega_{ij} \\ -\sin\omega_{ij} & \cos\omega_{ij} \end{bmatrix}.
	\end{equation}
	This submatrix is embedded in an $N \times N$ identity matrix, replacing columns $(j, j+1)$:
	\begin{equation}
	G_{R_{ij}} =
	\begin{bmatrix}
	1 & \cdots & 0 & 0 & 0 & \cdots & 0 \\
	\vdots & \ddots & \vdots & \vdots & \vdots & \ddots & \vdots \\
	0 & \cdots & 1 & 0 & 0 & \cdots & 0 \\
	0 & \cdots & 0 & e^{-i\theta_{ij}}\cos\omega_{ij} & e^{-i\theta_{ij}}\sin\omega_{ij} & \cdots & 0 \\
	0 & \cdots & 0 & -\sin\omega_{ij} & \cos\omega_{ij} & \cdots & 0 \\
	0 & \cdots & 0 & 0 & 0 & \cdots & 0 \\
	\vdots & \ddots & \vdots & \vdots & \vdots & \ddots & \vdots \\
	0 & \cdots & 0 & 0 & 0 & \cdots & 1
	\end{bmatrix}
	\end{equation}
	Right multiplication $U' = U \cdot G_{R_{ij}}$ mixes columns $j$ and $j+1$, zeroing $u'_{ij}$ while preserving unitarity $U'^{\dagger}U' = I$.
	
	The $L_{ij}$ primitive operates analogously via left multiplication, using element $u_{ij}$ and its upper neighbor $u_{i-1,j}$:
	\begin{align*}
	r &= u_{ij}/u_{i-1,j}, \quad \theta_{ij} = \arg(r), \quad \omega_{ij} = \arctan(-|r|).
	\end{align*}
	The $2 \times 2$ MZI transformation matrix is:
	\begin{equation}
	M_{L,ij} = \begin{bmatrix} e^{i\theta_{ij}}\cos\omega_{ij} & -\sin\omega_{ij} \\ e^{i\theta_{ij}}\sin\omega_{ij} & \cos\omega_{ij} \end{bmatrix}.
	\end{equation}
	This submatrix is embedded in an $N \times N$ identity matrix, replacing rows $(i-1, i)$:
	\begin{equation}
	G_{L_{ij}} =
	\begin{bmatrix}
	1 & \cdots & 0 & 0 & 0 & \cdots & 0 \\
	\vdots & \ddots & \vdots & \vdots & \vdots & \ddots & \vdots \\
	0 & \cdots & 1 & 0 & 0 & \cdots & 0 \\
	0 & \cdots & 0 & e^{i\theta_{ij}}\cos\omega_{ij} & -\sin\omega_{ij} & \cdots & 0 \\
	0 & \cdots & 0 & e^{i\theta_{ij}}\sin\omega_{ij} & \cos\omega_{ij} & \cdots & 0 \\
	0 & \cdots & 0 & 0 & 0 & \cdots & 0 \\
	\vdots & \ddots & \vdots & \vdots & \vdots & \ddots & \vdots \\
	0 & \cdots & 0 & 0 & 0 & \cdots & 1
	\end{bmatrix}.
	\end{equation}
	Left multiplication $U' = G_{L_{ij}} \cdot U$ mixes rows $i-1$ and $i$, zeroing $u'_{ij}$ while preserving unitarity $U'^{\dagger}U' = I$.
	
	All MZI primitives share the uniform type signature $\text{tmatrix} \to \text{tmatrix}$, so they can be freely chained into sequences. We present complete decompositions as ordered products of MZI primitives: each $L_{ij}$ left-multiplies the current matrix and each $R_{ij}$ right-multiplies it, so an ordered sequence such as $L_{4,1}$ followed by $R_{2,1}$ is written as $D=L_{4,1}\,U\,R_{2,1}$. The enumerative search explores different orderings and combinations of these primitives, constrained by type checking and guided by the learned grammar probabilities.
	
	Task generation creates unitary input matrices $\{U_k\}$ via QR decomposition of complex Gaussian matrices, with each task requiring the input to be transformed into diagonal form. Tasks support curriculum learning through a factory mechanism that applies a specified sequence of input transformation steps before presenting the matrix to the solver, enabling progressive learning of decomposition sub-problems.
	
	The search engine evaluates each candidate MZI sequence by checking two criteria: (1) whether all off-diagonal elements satisfy $|u_{ij}| < \epsilon_{\text{elem}}$ for $i \neq j$ (the hard diagonalization constraint), and (2) its posterior score $P(p|T)$ as defined above. Programs failing the hard constraint are immediately discarded. For programs passing it, an accuracy score is computed and combined with a simplicity prior (favoring shorter sequences) to form the posterior probability $P(p|T) \propto P(T|p)P(p)$. The search enumerates candidate sequences in order of decreasing prior probability under the learned grammar, so that concise and accurate decompositions are found first. Search depth limits are set according to matrix size to control the enumeration space.

	\subsection*{CURRICULUM LEARNING STRATEGY}
	
	For $5 \times 5$ matrices, independent learning successfully discovers decompositions with optimal MZI counts by constructing complete solutions from scratch. However, scaling to $6 \times 6$ and larger dimensions benefits from curriculum learning, a training strategy that progressively increases task difficulty. The curriculum consists of subtasks $\{T_k, T_{k-1}, ..., T_0\}$ of increasing complexity, where task $T_i$ provides partially decomposed matrices with the first $i$ steps of a reference elimination sequence applied purely to constrain the initial search space. For instance, $T_3$ presents matrices where the first 3 steps have already been executed, requiring the system to discover the remaining sequence. Early curriculum stages (large $i$) present tractable search spaces with more pre-applied steps and fewer remaining steps, enabling rapid discovery of short decomposition subsequences. The sleep phase abstracts these solutions into reusable library functions, which later stages leverage to construct longer programs through composition, avoiding exhaustive enumeration of deeply nested primitive sequences. This bootstrapping process is essential for $6 \times 6$ and larger matrices, where the lengthy search space proves intractable without hierarchical abstraction. Tasks are generated by applying a specified sequence of MZI transformations to randomly generated unitary matrices before presenting them to the solver. In our experiments, we use a curriculum depth of 3, training on tasks $\{T_3, T_2, T_1, T_0\}$ sequentially, progressing from simpler subtasks (fewer remaining steps) to more complex ones (more remaining steps). Deeper curricula can be used to discover decompositions for larger matrices directly, though this is not necessary since patterns learned from smaller matrices generalize compositionally. The wake-sleep cycle iterates within each curriculum stage until convergence criteria are met (e.g., solving all training instances or reaching iteration limits or timeout), then advances to the next stage.
	
	\subsection*{Results for $6 \times 6$ Matrices}
	
	Scaling to $6 \times 6$ matrices required curriculum learning, progressively training on subtasks of increasing difficulty to bootstrap library construction. After 7 iterations, the system discovered programs achieving the theoretical 15 MZI minimum. Table~\ref{tab:n6programs} lists five representative optimal decompositions discovered by the system.
	
\begin{figure}[h!]
		\centering
		\includegraphics[width=0.45\textwidth]{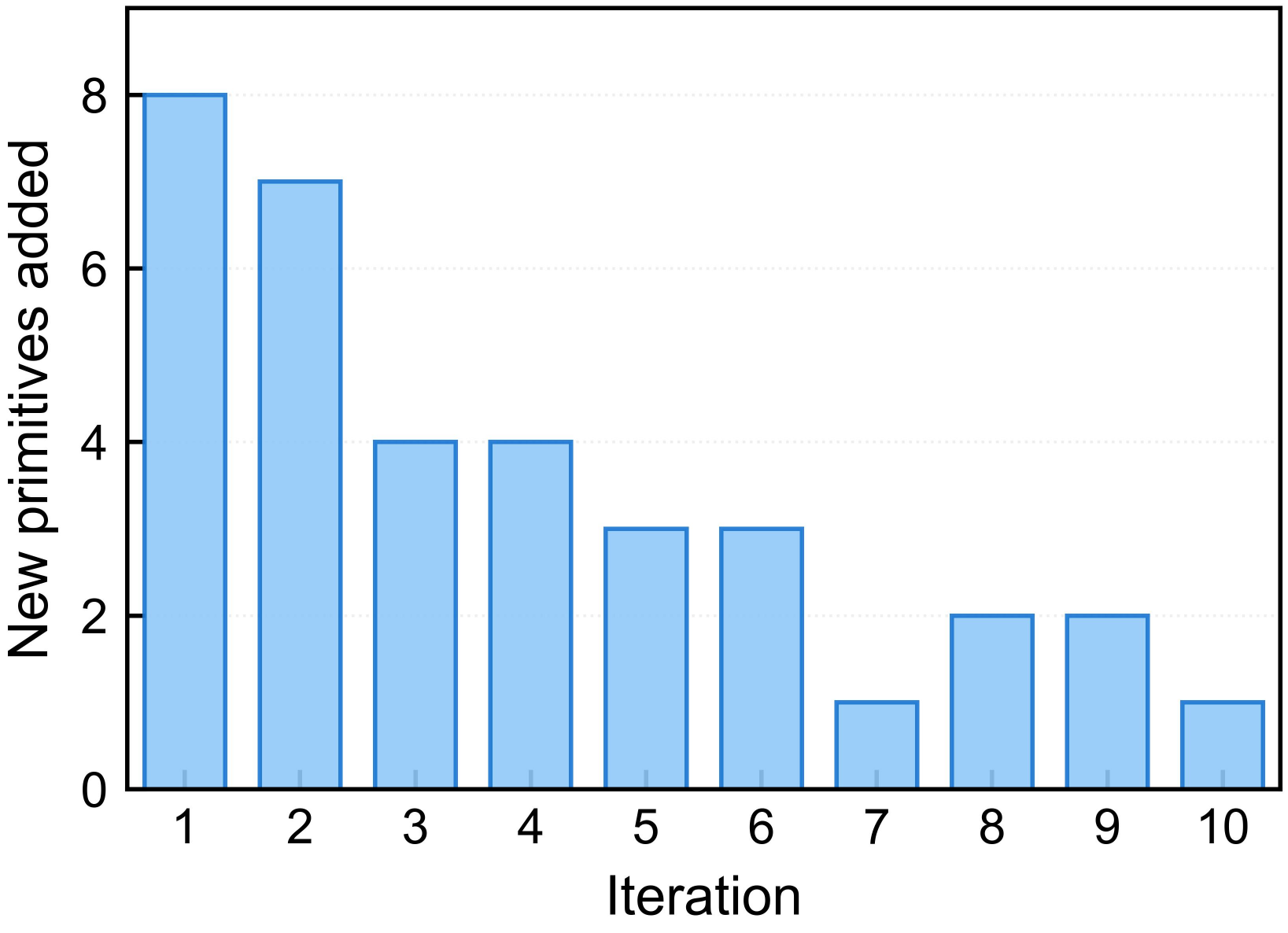}
		\caption{
			New primitives added per iteration during curriculum-assisted learning for $6 \times 6$ matrices.
		}
		\label{fig:delta_primitives}
	\end{figure}

	The evolution of the learned library reveals how the system builds hierarchical abstractions. Starting from 30 hand-defined MZI primitives, the grammar expanded to 65 primitives over 10 iterations via wake-sleep cycles. Fig.~\ref{fig:delta_primitives} shows rapid growth in early iterations with 8 and 7 new primitives in iterations 1-2, followed by stabilization with 1-2 new primitives per iteration after iteration 7, reflecting the transition from exploratory pattern discovery to consolidation of a compact, reusable vocabulary for expressing decompositions.
	
	This growth pattern explains curriculum learning's effectiveness: richer libraries enable reuse of extracted transformations, constructing deeper decomposition programs with less redundancy and revealing hierarchical organization in the learned vocabulary.

	\subsection*{REPRESENTATIVE SYNTHESIZED PROGRAMS}
	
	Tables \ref{tab:n5programs} and \ref{tab:n6programs} present representative decomposition programs discovered by the DreamCoder framework under independent learning and curriculum learning configurations, respectively. All programs are validated to successfully reduce randomly generated unitary matrices to diagonal form with all off-diagonal elements satisfying $|D_{ij}| < \epsilon_{\text{elem}}$ for $i \neq j$.	
	\begin{table}[htbp]
		\centering
		\caption{$5 \times 5$ programs (independent learning). All programs are validated on random unitary matrices.}
		\label{tab:n5programs}
		
		\begingroup
		\setlength{\tabcolsep}{10pt}
		\renewcommand{\arraystretch}{1.15}
		\large
		
		\begin{tabular}{lll}
			\hline\hline
			\# & Log-Posterior & Execution Order \\
			\hline
			1 & $-0.69$ &
			\parbox[t]{0.72\columnwidth}{\raggedright
				$\mathtt{L}_{40},\ \mathtt{R}_{41},\ \mathtt{R}_{42},\ \mathtt{R}_{43},\
				\mathtt{R}_{30},\ \mathtt{R}_{31},\ \mathtt{R}_{32},\ \mathtt{R}_{20},\
				\mathtt{R}_{21},\ \mathtt{R}_{10}$} \\
			2 & $-11.6$ &
			\parbox[t]{0.72\columnwidth}{\raggedright
				$\mathtt{L}_{40},\ \mathtt{L}_{30},\ \mathtt{R}_{41},\ \mathtt{R}_{42},\
				\mathtt{R}_{43},\ \mathtt{R}_{31},\ \mathtt{R}_{32},\ \mathtt{R}_{20},\
				\mathtt{R}_{21},\ \mathtt{R}_{10}$} \\
			3 & $-11.6$ &
			\parbox[t]{0.72\columnwidth}{\raggedright
				$\mathtt{L}_{40},\ \mathtt{R}_{41},\ \mathtt{R}_{42},\ \mathtt{L}_{30},\
				\mathtt{R}_{31},\ \mathtt{R}_{43},\ \mathtt{R}_{32},\ \mathtt{R}_{20},\
				\mathtt{R}_{21},\ \mathtt{R}_{10}$} \\
			4 & $-11.6$ &
			\parbox[t]{0.72\columnwidth}{\raggedright
				$\mathtt{L}_{40},\ \mathtt{R}_{41},\ \mathtt{R}_{30},\ \mathtt{R}_{42},\
				\mathtt{R}_{43},\ \mathtt{R}_{31},\ \mathtt{R}_{32},\ \mathtt{R}_{20},\
				\mathtt{R}_{21},\ \mathtt{R}_{10}$} \\
			5 & $-12.3$ &
			\parbox[t]{0.72\columnwidth}{\raggedright
				$\mathtt{L}_{40},\ \mathtt{L}_{30},\ \mathtt{L}_{41},\ \mathtt{R}_{42},\
				\mathtt{R}_{43},\ \mathtt{R}_{31},\ \mathtt{R}_{32},\ \mathtt{R}_{20},\
				\mathtt{R}_{21},\ \mathtt{R}_{10}$} \\
			\hline\hline
		\end{tabular}
		\endgroup
	\end{table}

	\begin{table}[htbp]
		\centering
		\caption{$6 \times 6$ programs (curriculum learning). All programs are validated on random unitary matrices.}
		\label{tab:n6programs}
		
		\begingroup
		\setlength{\tabcolsep}{10pt}
		\renewcommand{\arraystretch}{1.15}
		\large
		
		\begin{tabular}{lll}
			\hline\hline
			\# & Log-Posterior & Execution Order \\
			\hline
			1 & $-1.39$ &
			\parbox[t]{0.72\columnwidth}{\raggedright
				$\mathtt{R}_{50},\ \mathtt{R}_{51},\ \mathtt{R}_{52},\ \mathtt{R}_{53},\ \mathtt{R}_{54},\
				\mathtt{R}_{40},\ \mathtt{R}_{41},\ \mathtt{R}_{42},\ \mathtt{R}_{43},\
				\mathtt{R}_{30},\ \mathtt{R}_{31},\ \mathtt{R}_{20},\ \mathtt{R}_{32},\
				\mathtt{R}_{21},\ \mathtt{R}_{10}$} \\
			2 & $-1.39$ &
			\parbox[t]{0.72\columnwidth}{\raggedright
				$\mathtt{L}_{50},\ \mathtt{R}_{51},\ \mathtt{R}_{52},\ \mathtt{R}_{53},\ \mathtt{R}_{54},\
				\mathtt{R}_{40},\ \mathtt{R}_{41},\ \mathtt{R}_{42},\ \mathtt{R}_{43},\
				\mathtt{R}_{30},\ \mathtt{R}_{31},\ \mathtt{R}_{20},\ \mathtt{R}_{32},\
				\mathtt{R}_{21},\ \mathtt{R}_{10}$} \\
			3 & $-15.5$ &
			\parbox[t]{0.72\columnwidth}{\raggedright
				$\mathtt{R}_{50},\ \mathtt{L}_{40},\ \mathtt{R}_{51},\ \mathtt{R}_{52},\ \mathtt{R}_{53},\
				\mathtt{R}_{54},\ \mathtt{R}_{41},\ \mathtt{R}_{42},\ \mathtt{R}_{30},\
				\mathtt{R}_{43},\ \mathtt{R}_{31},\ \mathtt{R}_{20},\ \mathtt{R}_{32},\
				\mathtt{R}_{21},\ \mathtt{R}_{10}$} \\
			4 & $-20.1$ &
			\parbox[t]{0.72\columnwidth}{\raggedright
				$\mathtt{R}_{50},\ \mathtt{R}_{51},\ \mathtt{R}_{52},\ \mathtt{R}_{53},\ \mathtt{R}_{54},\
				\mathtt{L}_{40},\ \mathtt{R}_{41},\ \mathtt{R}_{42},\ \mathtt{R}_{43},\
				\mathtt{R}_{30},\ \mathtt{R}_{31},\ \mathtt{R}_{32},\ \mathtt{R}_{20},\
				\mathtt{R}_{21},\ \mathtt{R}_{10}$} \\
			5 & $-23.1$ &
			\parbox[t]{0.72\columnwidth}{\raggedright
				$\mathtt{L}_{50},\ \mathtt{L}_{40},\ \mathtt{L}_{51},\ \mathtt{R}_{52},\ \mathtt{R}_{53},\
				\mathtt{R}_{54},\ \mathtt{R}_{41},\ \mathtt{R}_{42},\ \mathtt{R}_{30},\
				\mathtt{R}_{43},\ \mathtt{R}_{31},\ \mathtt{R}_{20},\ \mathtt{R}_{32},\
				\mathtt{R}_{21},\ \mathtt{R}_{10}$} \\
			\hline\hline
		\end{tabular}
		\endgroup
	\end{table}

	\subsection*{STRUCTURE-AWARE DECOMPOSITIONS}
	
	When the input matrix carries structural constraints---such as Householder form or sparsity-induced structural constraints---the synthesized programs automatically exploit them, discovering far fewer MZI operations than the general-case bound. We present two families of examples.
	
	\textbf{Householder reflectors.} A Householder matrix $H = I - 2\mathbf{v}\mathbf{v}^\dagger$ (where $\mathbf{v} \in \mathbb{C}^N$, $\|\mathbf{v}\|=1$) can be decomposed using only $2N-3$ MZI operations, compared to the $N(N-1)/2$ required for a general unitary. The synthesis system discovers programs achieving this reduction without being provided with the analytical formula of Householder reflectors.
	
	For $N=5$, every Householder task is solved in exactly 7 operations ($=2\times5-3$), with the representative program $\mathtt{L}_{4,0},\mathtt{R}_{4,3},\mathtt{R}_{3,0},\mathtt{R}_{3,1},\mathtt{R}_{3,2},\mathtt{R}_{2,1},\mathtt{R}_{1,0}$. For $N=6$, every task is solved in exactly 9 operations ($=2\times6-3$), with the representative program $\mathtt{L}_{5,0},\mathtt{R}_{5,4},\mathtt{L}_{4,0},\mathtt{R}_{4,3},\mathtt{R}_{3,0},\mathtt{R}_{3,1},\mathtt{R}_{3,2},\mathtt{R}_{2,1},\mathtt{R}_{1,0}$. The $N\to N+1$ extension prepends $\mathtt{L}_{N,0}$ and $\mathtt{R}_{N,N-1}$ as the innermost pair, consistent with $2N-3$ scaling.
	
	\textbf{Sparse SVD decompositions.} When the source matrix $W$ is sparse, its SVD $W=U\Sigma V^\dagger$ yields unitary factors $U$ and $V^\dagger$ with internal structure that reduces the number of MZI operations needed for decomposition. The system discovers and exploits this structure automatically from examples, without being provided with information about the sparsity level or structure of the input.
	
	\textit{NNZ-controlled sparsity.} We consider a representative example with source $W$: $81\%$ sparse, $7/36$ nonzero entries controlled by NNZ. The input $U$ matrix is shown below to three decimal places for readability, with all computations performed at full floating-point precision.
	\[
	U = \begin{pmatrix}
	{-}0.686{-}0.039i & {-}0.696{-}0.039i &  0.000{+}0.000i &  0.204{+}0.011i &  0.000{+}0.000i &  0.000{+}0.000i \\
	0.000{+}0.000i &  0.000{+}0.000i &  0.444{-}0.896i &  0.000{+}0.000i &  0.000{+}0.000i &  0.000{+}0.000i \\
	{-}0.184{-}0.078i &  0.416{+}0.177i &  0.000{+}0.000i &  0.800{+}0.340i &  0.000{+}0.000i &  0.000{+}0.000i \\
	0.000{+}0.000i &  0.000{+}0.000i &  0.000{+}0.000i &  0.000{+}0.000i &  0.896{-}0.443i &  0.000{+}0.000i \\
	0.000{+}0.000i &  0.000{+}0.000i &  0.000{+}0.000i &  0.000{+}0.000i &  0.000{+}0.000i &  0.251{+}0.968i \\
	{-}0.400{-}0.572i &  0.319{+}0.456i &  0.000{+}0.000i & {-}0.258{-}0.369i &  0.000{+}0.000i &  0.000{+}0.000i
	\end{pmatrix}.
	\]
	The corresponding diagonalization relation is
	\[
	D = L_{3,0}L_{4,0}L_{5,0}\,U\,R_{3,1}R_{3,2}R_{2,0}R_{2,1}R_{1,0},
	\]
	achieving a $47\%$ reduction from the universal theoretical bound of 15 MZIs, verified at machine precision ($\sim10^{-16}$).
	
	\textit{Bernoulli sparsity.} We consider a representative example with source $W$: $86.1\%$ sparse, $5/36$ nonzero; each entry independently zeroed with probability $80\%$. The input $U$ matrix is
	\[
	U = \begin{pmatrix}
	0.000{+}0.000i & 0.000{+}0.000i & 1.000{+}0.000i & 0.000{+}0.000i & 0.000{+}0.000i & 0.000{+}0.000i \\
	0.336{+}0.000i & 0.942{+}0.000i & 0.000{+}0.000i & 0.000{+}0.000i & 0.000{+}0.000i & 0.000{+}0.000i \\
	{-}0.532{-}0.643i & 0.190{+}0.229i & 0.000{+}0.000i & 0.173{-}0.430i & 0.000{+}0.000i & 0.000{+}0.000i \\
	0.000{+}0.000i & 0.000{+}0.000i & 0.000{+}0.000i & 0.000{+}0.000i & {-}1.000{+}0.015i & 0.000{+}0.000i \\
	0.000{+}0.000i & 0.000{+}0.000i & 0.000{+}0.000i & 0.000{+}0.000i & 0.000{+}0.000i & 0.970{+}0.243i \\
	0.111{-}0.422i & {-}0.040{+}0.151i & 0.000{+}0.000i & {-}0.861{+}0.212i & 0.000{+}0.000i & 0.000{+}0.000i
	\end{pmatrix}.
	\]
	The corresponding diagonalization relation is
	\[
	D = L_{4,0}L_{5,0}\,U\,R_{3,1}R_{1,0}R_{2,1}R_{3,2},
	\]
	achieving a 60\% reduction from the universal theoretical bound of 15 MZIs, verified at machine precision ($\sim10^{-16}$).
	
	Across both structure classes, the synthesis system discovers these reductions automatically, without being provided with the analytical forms or sparsity properties of the input matrices, demonstrating that structure-aware efficiency emerges naturally from the program synthesis process itself.
	
	\subsection*{CROSS-DIMENSIONAL GENERALIZATION: EXTRACTING DIMENSION-INDEPENDENT CONSTRUCTION RULES FROM SYNTHESIZED PROGRAMS}
	
	A critical capability of the discovered decomposition programs is their ability to generalize across matrix dimensions without retraining. The key mechanism is that the synthesis system does not memorize a fixed operation sequence at low $N$, but extracts a dimension-independent construction rule. Once such a rule is identified, substituting any target $N$ directly generates the corresponding execution sequence without retraining. We illustrate this with two concrete examples before presenting the detailed pseudocode for Pattern~4.
	
	\textbf{How low-dimensional patterns are expanded to high dimensions.}
	
	\textit{Example 1 --- Sequential row elimination (Table~\ref{tab:n5programs}, Program~1):} the $N=5$ sequence is $\mathtt{L}_{4,0}, \mathtt{R}_{4,1}, \mathtt{R}_{4,2}, \mathtt{R}_{4,3}, \mathtt{R}_{3,0}, \mathtt{R}_{3,1}, \mathtt{R}_{3,2}, \mathtt{R}_{2,0}, \mathtt{R}_{2,1}, \mathtt{R}_{1,0}$. The rule is: for the top row ($i=N-1$), apply $\mathtt{L}_{i,0}$ followed by $\mathtt{R}_{i,1}, \ldots, \mathtt{R}_{i,i-1}$. For all lower rows ($i = N-2$ down to $1$), apply $\mathtt{R}_{i,0}, \mathtt{R}_{i,1}, \ldots, \mathtt{R}_{i,i-1}$ in order. Each row is fully eliminated before moving to the next. The operation count follows directly: $\sum_{i=1}^{N-1} i = N(N-1)/2$, yielding 2016 operations at $N=64$.
	
	\textit{Example 2 --- Two-row leading elimination (Table~\ref{tab:n5programs}, Program~2):} the $N=5$ sequence is $\mathtt{L}_{4,0}, \mathtt{L}_{3,0}, \mathtt{R}_{4,1}, \mathtt{R}_{4,2}, \mathtt{R}_{4,3}, \mathtt{R}_{3,1}, \mathtt{R}_{3,2}, \mathtt{R}_{2,0}, \mathtt{R}_{2,1}, \mathtt{R}_{1,0}$. The rule is: first apply $\mathtt{L}_{N-1,0}$ and $\mathtt{L}_{N-2,0}$; then complete row $N-1$ with $\mathtt{R}_{N-1,1}, \ldots, \mathtt{R}_{N-1,N-2}$; then row $N-2$ with $\mathtt{R}_{N-2,1}, \ldots, \mathtt{R}_{N-2,N-3}$ (starting from $j=1$ since $j=0$ was handled by $\mathtt{L}_{N-2,0}$); then all remaining rows $i = N-3$ down to $1$ with $\mathtt{R}_{i,0}, \mathtt{R}_{i,1}, \ldots, \mathtt{R}_{i,i-1}$. The same operation count $N(N-1)/2$ is preserved.
	
	We verified Programs 1, 2, and 3 from Table~\ref{tab:n5programs} across $N = 5, 6, 7, 8, 16, 32, 64$; all achieve machine-precision accuracy ($\sim 10^{-16}$) with 100\% success rate, as shown in Table~\ref{tab:generalization_precision}. This confirms that the decompositions are mathematically exact: the residual error reflects only floating-point rounding and does not grow systematically with $N$.
	
	\begin{table}[htbp]
		\centering
		\caption{Numerical precision verification for three universal programs across different dimensions. Off-diagonal error averaged over 5 random unitary matrices.}
		\label{tab:generalization_precision}
		\begin{tabular}{cccc}
			\hline
			$N$ & Program 1 & Program 2 & Program 3 \\
			\hline
			5  & $2.61 \times 10^{-16}$ & $2.92 \times 10^{-16}$ & $2.73 \times 10^{-16}$ \\
			6  & $3.10 \times 10^{-16}$ & $3.32 \times 10^{-16}$ & $3.12 \times 10^{-16}$ \\
			7  & $4.38 \times 10^{-16}$ & $4.65 \times 10^{-16}$ & $4.23 \times 10^{-16}$ \\
			8  & $4.20 \times 10^{-16}$ & $3.92 \times 10^{-16}$ & $4.66 \times 10^{-16}$ \\
			16 & $4.98 \times 10^{-16}$ & $4.73 \times 10^{-16}$ & $4.67 \times 10^{-16}$ \\
			32 & $4.71 \times 10^{-16}$ & $4.84 \times 10^{-16}$ & $5.38 \times 10^{-16}$ \\
			64 & $5.69 \times 10^{-16}$ & $6.04 \times 10^{-16}$ & $6.24 \times 10^{-16}$ \\
			\hline
		\end{tabular}
	\end{table}
	
	We next present the detailed algorithmic procedure for Pattern~4 (row-pair interleaving, Fig.~\ref{fig:n5patterns}; distinct from the programs listed in Table~\ref{tab:n5programs}) as a representative example, with explicit pseudocode for constructing the decomposition sequence at arbitrary $N$.
	
	The learned decomposition strategies encode systematic elimination orderings that can be expressed through regular patterns in the primitive sequence. Specifically, Pattern~4 implements a \emph{row-pair interleaving} strategy: starting from the largest row index $i = N-1$ and decrementing by two each step, the algorithm processes rows in pairs by first emitting primitive $\mathtt{R}_{i,0}$, then interleaving primitives from two adjacent rows as $\mathtt{R}_{i,j} \to \mathtt{R}_{i-1,\,j-1}$ for $j = 1, \ldots, i-1$. When a single row remains ($i=1$), $\mathtt{R}_{1,0}$ closes the sequence. This fixed rule enumerates all $N(N-1)/2$ operations exactly and generalizes unchanged to arbitrary dimensions.
	
	The following pseudocode formalizes this generalization procedure for constructing lambda expressions representing $N \times N$ unitary decompositions:
	
	\begin{verbatim}
	Algorithm: GenerateDimensionAgnosticDecomposition(N)
	Input: Matrix dimension N
	Output: Lambda expression for NxN unitary decomposition
	
	1: Initialize operation_sequence := []
	2: i := N - 1  // Start from the largest row index
	3:
	4: while i >= 1 do
	5:     if i = 1 then
	6:         // Base case: single remaining row
	7:         operation_sequence.append((1, 0))
	8:         break
	9:     end if
	10:
	11:    // Process row pair (i, i-1) with interleaving
	12:    // First emit the leading element of row i
	13:    operation_sequence.append((i, 0))
	14:
	15:    // Then interleave elements from rows i and i-1
	16:    for j = 1 to i-1 do
	17:        operation_sequence.append((i, j))      // Row i element
	18:        operation_sequence.append((i-1, j-1))  // Row i-1 element
	19:    end for
	20:
	21:    i := i - 2  // Move to next row pair
	22: end while
	23:
	24: // Construct lambda expression from innermost to outermost
	25: lambda_expr := "$0"  // Input matrix placeholder
	26:
	27: for each (row, col) in operation_sequence do
	28:     // Wrap with primitive R_{row,col}
	29:     lambda_expr := "(R_{" + row + col + "} " + lambda_expr + ")"
	30: end for
	31:
	32: lambda_expr := "(lambda " + lambda_expr + ")"
	33: return lambda_expr
	\end{verbatim}
	
	This construction ensures that the total number of primitives equals $N(N-1)/2$, the theoretical minimum for universal unitary decomposition. The dimension-agnostic nature of this pattern arises from its reliance on structural invariants—specifically, the row-pair interleaving rule and the fixed progression from larger to smaller row indices—rather than dimension-specific heuristics. Consequently, the same algorithmic template applies to matrices of any size, enabling automatic scaling from training dimensions (e.g., $5 \times 5$) to deployment dimensions (e.g., $64 \times 64$ or larger).
	
	The generalization validation protocol tests learned programs on matrices of progressively increasing dimensions: $8 \times 8$, $16 \times 16$, $32 \times 32$, and $64 \times 64$. For each test dimension, we verify that the compositionally extended program achieves successful diagonalization on random unitary matrices with all off-diagonal elements satisfying $|p(U_k)_{ij}| < \epsilon_{\text{elem}}$ for all $k$ and all $i \neq j$, with residuals at machine precision ($\sim 10^{-16}$). Successful verification across all test dimensions confirms that the discovered decomposition programs encode dimension-agnostic decomposition strategies rather than memorized dimension-specific patterns, establishing the foundation for scalable photonic circuit synthesis.

\end{document}